\documentclass[aps, preprint, groupedaddress, floatfix, onecolumn]{revtex4-2}

\usepackage{siunitx}
\usepackage{booktabs}
\usepackage{chemformula}
\let\ce\ch
\usepackage{orcidlink}
\usepackage{lineno}
\usepackage{xcolor}
\usepackage[T1]{fontenc}
\usepackage{bm}
\usepackage{float}

\usepackage{hyperref}

\newcommand{\kl}{${\kappa_{\rm L}}$}

\newcolumntype{Y}{>{\centering\arraybackslash}X}
\newcolumntype{Z}{>{\hsize=1.1\hsize\centering\arraybackslash}X}

\begin{document}

\title{Ultralow thermal conductivity via weak interactions in PbSe/PbTe monolayer heterostructure for thermoelectric design}

\author{Ruihao Tan\orcidlink{0009-0000-1145-9448}${^1}$}
\author{Kaiwang Zhang\orcidlink{0000-0002-3538-7001}${^1}$}
\email{kwzhang@xtu.edu.cn}
\author{Yue-Wen Fang\orcidlink{0000-0003-3674-7352}${^2}$}
\email{yuewen.fang@ehu.eus}
\affiliation{     $^1$ School of Physics and Optoelectronics, Xiangtan University, Xiangtan, 411105, Hunan, China.}
\affiliation{     $^2$ Centro de F{\'i}sica de Materiales (CFM-MPC), CSIC-UPV/EHU, Manuel de Lardizabal Pasealekua 5, 20018 Donostia/San Sebasti{\'a}n, Spain.}

\begin{abstract}
In this study, we systematically investigate the thermal and electronic transport properties of two-dimensional PbSe/PbTe monolayer heterostructure by combining first-principles calculations, Boltzmann transport theory, and machine learning methods. The heterostructure exhibits a unique honeycomb-like corrugated and asymmetric configuration, which significantly enhances phonon scattering. Moreover, the relatively weak interatomic interactions in PbSe/PbTe lead to the formation of anti-bonding states, resulting in strong anharmonicity and ultimately yielding ultralow lattice thermal conductivity (\kl). In the four-phonon scattering model, the \kl~values along the $x$ and $y$ directions are as low as 0.37 and 0.31 W/mK, respectively. Contrary to the conventional view that long mean free path acoustic phonons dominate heat transport, we find that optical phonons contribute approximately 59\% of the lattice thermal conductivity in this heterostructure. These optical phonons exhibit large Grüneisen parameters, strong anharmonic scattering, and relatively high group velocities, thereby playing a crucial role in the low \kl~regime. Further analysis of thermoelectric performance shows that at a high temperature of 800 K, the heterostructure achieves an exceptional dimensionless figure of merit ($ZT$) of 5.3 along the $y$ direction, indicating outstanding thermoelectric conversion efficiency. These findings not only provide theoretical insights into the transport mechanisms of PbSe/PbTe monolayer heterostructure but also offer a practical design strategy for developing high-performance two-dimensional layered thermoelectric materials.

\end{abstract}
\maketitle
\clearpage

\section{Introduction}

Driven by the record-high global temperatures and the increased industry consumption, the global energy demand in 2024 rose by \SI{2.2}{\percent}, according to \textit{Global Energy Review 2025} by the International Energy Agency. This growth rate is considerably higher than the average annual increase of \SI{1.3}{\percent} between 2013 and 2023. The emerging energy crisis has gained global attention and has intensified scientific pursuit of advanced materials capable of providing efficient and sustainable energy solutions. As the energy crisis intensifies, thermoelectric materials are receiving more attention than ever before due to their ability of converting waste heat into clean energy through the direct and reversible conversion between heat and electricity via the Seebeck and Peltier effects~\cite{Qin2024,Liu2025,Hu2021Adv,Hu2021Nat}. The conversion efficiency is typically evaluated using the dimensionless figure of merit:

\begin{equation}
ZT = \frac{S^2 \sigma T}{\kappa_e + \kappa_L}
\end{equation}

where $S$ is the Seebeck coefficient, $\sigma$ is the electrical conductivity, $T$ is the absolute temperature, and $\kappa_e$ and $\kappa_L$ denote the electronic and lattice thermal conductivities, respectively. These parameters are inherently coupled in condensed matter systems, where the Seebeck coefficient $S$ typically exhibits an inverse relationship with electrical conductivity, and $\kappa_e$ is directly proportional to $\sigma$. Such interdependencies pose a significant challenge to optimizing $ZT$, as improving one parameter often compromises another~\cite{Qin2024,Hu2021Nat,ref5,ref6}. Several approaches including defect engineering~\cite{ref7,ref8,ref9,ref10}, band structure engineering~\cite{ref11,ref12}, doping strategies~\cite{Hu2021Adv,ref13,ref14}, and the construction of heterostructures~\cite{ref15,ref16} have been proposed to tune the electronic band structure or phonon transport, thereby improving thermoelectric figure of merit.

In the field of thermoelectrics, 2D materials exhibit unique advantages owing to quantum confinement effects and tunable interfacial phonon scattering. Their atomic-scale thickness can significantly suppress phonon transport while maintaining high carrier mobility~\cite{ref17,ref18}. Lee \textit{et al.} demonstrated that in conventional bulk materials, electrical conductivity ($\sigma$) and electronic thermal conductivity ($\kappa_e$) are positively correlated and difficult to decouple. However, in \ce{SnS2} nanosheets, reducing the thickness significantly lowers the lattice thermal conductivity ($\kappa_L$) while maintaining a moderate $\sigma$, ultimately enhancing thermoelectric efficiency~\cite{ref19}. This observation highlights the unique potential of 2D materials in the synergistic regulation of the power factor and the total thermal conductivity ($\kappa_L + \kappa_e$). In recent years, particular attention has been drawn to two-dimensional (2D) heterostructures with van der Waals (vdW) interactions, as the inherent asymmetry within these systems can enhance the coupling between acoustic and optical phonon modes, thereby strengthening interfacial phonon scattering and resulting in a reduced $\kappa_L$. These systems are typically constructed by stacking different 2D layers vertically through weak vdW forces. Alhemaitijiang Sidis \textit{et al.} employed interfacial engineering techniques to investigate black phosphorus/arsenic heterostructure devices, focusing on their thermoelectric behavior~\cite{ref20}. Their findings revealed a high figure of merit $ZT$ of up to 3.5 at \SI{350}{\kelvin}, highlighting the exceptional thermoelectric performance of such hybrid systems. In another study, Zhou \textit{et al.} investigated the in-plane thermoelectric properties of vertically stacked black phosphorus (\ce{BP}) and \ce{Ti2C}, forming van der Waals heterostructures (vdWHs). Remarkably, the $ZT$ value of the \ce{Ti2C/BP} vdWHs at room temperature (\SI{300}{\kelvin}) exhibited a remarkable improvement by approximately $10^3$ and $10^4$ times compared to pristine \ce{Ti2C} and \ce{BP}, respectively. This dramatic enhancement is attributed to the strong interlayer coupling, which significantly boosts thermoelectric efficiency and positions these vdWHs as promising candidates for next-generation thermoelectric technologies~\cite{ref21}. Wu \textit{et al.} also reported significant progress in this field, demonstrating a remarkable improvement in the thermoelectric performance of \ce{MoS2/h-BN} heterostructures compared to their monolayer counterparts. In particular, the thermoelectric power factor of the \ce{MoS2/h-BN} device exhibited an enhancement of two orders of magnitude over that of monolayer \ce{MoS2}, validating the effectiveness of this integration strategy for energy conversion applications~\cite{ref22}. Additionally, research by Tang \textit{et al.} demonstrated that the \ce{RbSe/SnSe} heterostructure achieved a $ZT$ value of 2.6 at \SI{900}{\kelvin}, representing a remarkable improvement of approximately \SI{100}{\percent} and \SI{13}{\percent} over the individual \ce{RbSe} ($ZT=1.3$) and \ce{SnSe} ($ZT=2.3$) monolayers, respectively. This significant enhancement highlights the strong potential of such heterostructures for high-temperature thermoelectric applications~\cite{ref16}.

Recently, the monolayers \ce{RbTe} and \ce{RbSe} with same honeycomb-like wrinkled structures have been theoretically reported to show high $ZT$ values of 1.55 and 1.33 at \SI{900}{\kelvin}, respectively, based on the three-phonon scattering model~\cite{ref23,ref24}. In spite of the high thermoelectric performance of the individual monolayers, the structure and the thermoelectric properties of a heterostructure system consisting of monolayers \ce{RbTe} and \ce{RbSe} remain to be explored. Herein, we employ a combination of first-principles calculations, the Boltzmann transport equation, and machine learning algorithms to systematically investigate the crystal structure, electronic transport, phonon behavior, and thermoelectric performance of the wrinkled \ce{RbSe/RbTe} monolayer heterostructure. Through detailed analysis of interatomic bonding characteristics and anharmonic vibrational effects, we reveal the underlying physical mechanisms responsible for the intrinsically low lattice thermal conductivity of this type of heterostructure. By incorporating four-phonon scattering processes, our results show that the \ce{RbSe/RbTe} structure achieves maximum $ZT$ values of 4.1 and 5.3 along the $x$- and $y$-directions, respectively, at \SI{800}{\kelvin}. This work not only deepens the understanding of thermoelectric transport in such heterostructures, but also provides theoretical guidance and a practical design strategy for developing high-performance 2D layered thermoelectric materials.

\section{Methods}
The first-principles calculations based on density functional theory (DFT) were performed by using the VASP code~\cite{Kresse-PRB-1996,Kresse1996}, in which the projector augmented wave (PAW) approach were used for the accurate treatment of core valence interactions~\cite{Blochl1994,Kresse-PAW-PRB1999}. The exchange-correlation interactions within the PbSe/PbTe monolayer heterostructure were treated using the Perdew-Burke-Ernzerhof (PBE) functional in the generalized gradient approximation (GGA)~\cite{ref29}, A plane-wave kinetic energy cutoff of 500 eV was employed, and the Brillouin zone was sampled with a 12 $\times$ 12 $\times$ 1 Monkhorst–Pack \textbf{k}-point grid~\cite{Monkhorst-PRB1976}. To eliminate spurious interactions between periodic images, a vacuum region of 20 \AA~ was inserted along the out-of-plane direction. Convergence thresholds were set to 10$^{-8}$ eV for total energy and 0.001 eV/\AA~ for atomic forces. To accurately assess the electronic band structure, the Heyd–Scuseria–Ernzerhof (HSE06) hybrid functional was utilized~\cite{HSE06-JCP2003}. Thermal stability of the two-dimensional PbSe/PbTe monolayer heterostructure was verified through ab initio molecular dynamics (AIMD) simulations. Using a 4 $\times$ 4 $\times$ 1 supercell, the monolayer heterostructure was simulated at 300 K and 700 K for a total duration of 10 ps with a time step of 1 fs, employing the NVT ensemble. Additionally, crystal orbital Hamilton population (COHP) analysis was carried out using the Local Orbital Basis Suite Towards Electronic-Structure Reconstruction (LOBSTER) code~\cite{ref32}. To account for van der Waals interactions, the VASP calculations were performed with the IVDW = 11 parameter.

MLIP (Machine Learning Interatomic Potential) family offers high precision in modeling atomic interactions and is particularly compatible with the Moment Tensor Potential (MTP) approach. MTP is based on the locality principle, where the total energy of a system with $N$ atoms is described as a sum over atomic neighborhood contributions:
\begin{equation}
E = E^{\mathrm{MTP}} = \sum_i V(u_i) .
\end{equation}
Here, $V(u_i)$ represents the energy contribution from the neighborhood of atom $i$. Two atoms are considered neighbors if their separation is less than a defined cutoff distance $R_{\rm cut}$. In our case, $R_{\rm cut}$ = 5 \AA~was used. The neighborhood $u_i$ of atom $i$ includes the positions and types of all its neighbors~\cite{ref33}:
\begin{equation}
u_i = \left( \{ r_{i1}, Z_i, Z_1 \}, \ldots, \{ r_{ij}, Z_i, Z_j \}, \ldots, \{ r_{i N_{\text{neigh}}}, Z_i, Z_{N_{\text{neigh}}} \} \right),
\end{equation}
where \( r_{ij} \), \( z_i \), \( z_j \), and \( N_{\text{neigh}} \) denote the vector between atoms, the type of the central atom, the type of neighboring atoms, and the count of neighboring atoms, respectively. For a given atomic configuration, its energy contribution is represented as~\cite{ref33}:
${
V(u_i) = \sum_{\alpha} \xi_{\alpha} B_{\alpha}(u_i)
}$
where \( B_{\alpha} \) are the basis functions and \( \xi_{\alpha} \) are tunable parameters within the MLIP. A scalar quantity is then constructed by combining all possible contractions of moment tensor descriptors through these basis functions. A scalar based on basis functions constructed from all possible contractions of moment tensor descriptors is generated as follows~\cite{ref33}:
\begin{equation}
M_{\mu, \nu}(u) = \sum_{j=1}^{N_i^{\text{neigh}}} f_{\mu}(|\mathbf{r}_{ij}|, z_i, z_j) \left( \mathbf{r}_{ij} \right)^{\otimes \nu}.
\end{equation}
Here, the radial function \( f_{\mu} \) depends on the interatomic distance and atomic types, and is given by~\cite{ref33}:
\begin{equation}
f_{\mu}(|\mathbf{r}_{ij}|, z_i, z_j) = \sum_{\beta} c^{(z_i, z_j)}_{\mu \beta} \, \phi_{\beta}(|\mathbf{r}_{ij}|) \, \chi(R_{\text{cut}} - |\mathbf{r}_{ij}|).
\end{equation}
In this formulation, \( \phi_{\beta} \) are radial basis functions, \( \chi \) is a cutoff smoothing function, and \( c_{\mu \beta} \) are learnable coefficients.

To construct the MLIP, a MTP framework was adopted, where model fitting was carried out by minimizing a custom-designed loss function~\cite{Shapeev-MLIP-MultiModeling2016,Novikov_2021-MLIP-MLST}:
\begin{equation}
\sum_{k=1} \left[
    w_e \left( E_k^{\mathrm{AIMD}} - E_k^{\mathrm{MTP}} \right)
    + w_f \sum_{i}^{N} \left| \mathbf{f}_{k,i}^{\mathrm{AIMD}} - \mathbf{f}_{k,i}^{\mathrm{MTP}} \right|^2
    + w_s \sum_{i,j=1}^{3} \left| \sigma_{k,ij}^{\mathrm{AIMD}} - \sigma_{k,ij}^{\mathrm{MTP}} \right|^2
\right] \rightarrow \min .
\end{equation}
This objective function aims to align MTP predictions with reference data obtained from ab initio molecular dynamics (AIMD), including total energies, atomic forces, and stress tensors. Specifically, \( E_k^{\mathrm{AIMD}} \), \( \mathbf{f}_{k,i}^{\mathrm{AIMD}} \), and \( \sigma_{k,ij}^{\mathrm{AIMD}} \) represent the AIMD-obtained values, while \( E_k^{\mathrm{MTP}} \), \( \mathbf{f}_{k,i}^{\mathrm{MTP}} \), and \( \sigma_{k,ij}^{\mathrm{MTP}} \) are their MTP-computed counterparts. The non-negative importance weights of energies, forces, and stresses are represented by the symbols \( w_e \), \( w_f \), and \( w_s \), which are set to 1, 0.1, and 0.001, respectively. The dataset used for training was generated from AIMD simulations of a \( 4 \times 4 \times 1 \) PbSe/PbTe monolayer supercell. These AIMD simulations were performed in the canonical (NVT) ensemble at four temperatures: 50, 300, 500, and 700 K, employing a timestep of 1 fs and a total duration time of 1 ps for each trajectory. Model training was subsequently executed using the MLIP software suite~\cite{Novikov_2021-MLIP-MLST}.
Phonon calculations were calculated by combining the MTP potentials with the PHONOPY code~\cite{Togo-phonopy2015}, which enabled computation of phonon dispersion curves and second-order harmonic interatomic force constants (IFCs). For anharmonic interactions, third-order IFCs were obtained using the THIRDORDER.PY utility in conjunction with MTP, taking into account interactions extending to the eighth-nearest neighbors within an enlarged 6$\times$6$\times$1 simulation cell. Furthermore, fourth-order IFCs were calculated by interfacing the MTP framework with the FOURTHORDER.PY script~\cite{Fourphonon-Han-CPC2022}.
Based on the harmonic and higher-order IFCs, phonon Boltzmann transport equation (BTE) calculations were conducted to determine the lattice thermal conductivity. The ShengBTE and Fourphonon solvers were employed to include three-phonon and four-phonon scattering processes, respectively, utilizing dense \textbf{q}-point meshes of 60$\times$60$\times$1 and 35$\times$35$\times$1 to ensure convergence~\cite{ref33,ref38}. Supplementary Figure 1 presents the details of the \textbf{q}-mesh convergence test considering only the three-phonon scattering model, while Supplementary Figure 2 shows the details of the \textbf{q}-mesh convergence test considering both the three-phonon and four-phonon scattering models. The convergence criterion is set to 0.1 W/mK~\cite{ref39}. As for charge transport, key electronic properties, namely, the Seebeck coefficient ($S$), electrical conductivity ($\sigma$), and electronic thermal conductivity ($\kappa_{\rm e}$), were computed using the BoltzTraP code~\cite{ref40}. The carrier relaxation times were incorporated via deformation potential theory, and the fitting procedure is illustrated in Supplementary Figure 3.

\section{Results}
\subsection{Crystal Structure, Chemical Bonding, and Stability}
\begin{figure}[htp]
\centering
\includegraphics[angle=0,width=0.99\textwidth]{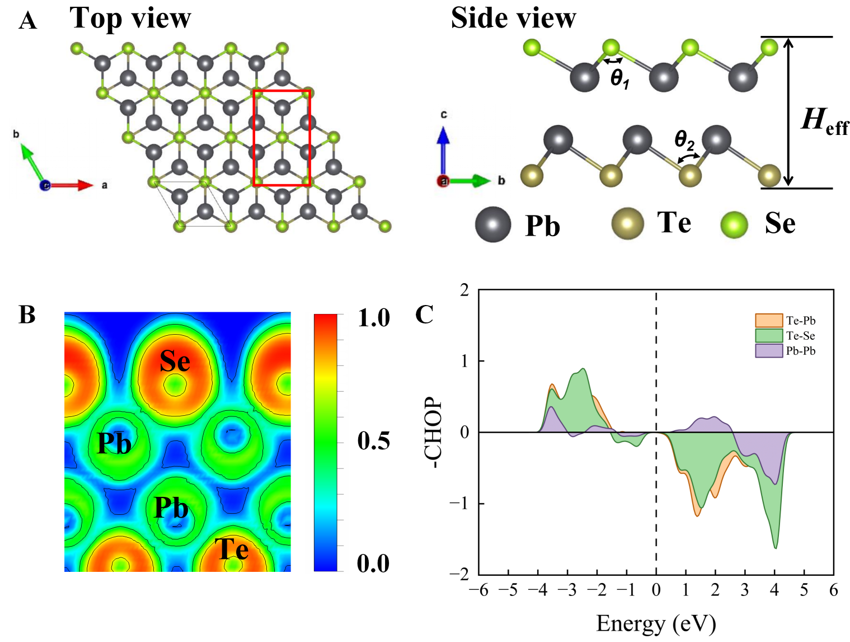}
\caption{\label{fig:structure-elf-cohp}  
Supercell structure of the PbSe/PbTe monolayer heterostructure: (A) Top view and side view; (B) Electron localization function (ELF); (C) -COHP analysis.
}
\end{figure}

As illustrated in Figure~\ref{fig:structure-elf-cohp}A, the monolayer PbSe/PbTe monolayer heterostructure exhibits a pronounced wrinkled morphology. The crystal structure consists of two Pb atomic layers sandwiched by alternately arranged Se and Te atoms, and belongs to the $P3m1$ (No. 156) space group. According to the structural parameters listed in Table~\ref{tab:lattice_params}, the in-plane lattice constant of the heterostructure is 4.30 \AA, which lies between those of monolayer PbSe (4.091 \AA) and PbTe (4.364 \AA). The atomic size mismatch between Te (atomic radius: 2.07 \AA) and Se (1.90 \AA) leads to noticeable asymmetry in the bond lengths and bond angles of the Pb–Te and Pb–Se bonds, resulting in local structural distortion at the heterointerface and the formation of a complex wrinkled configuration. This asymmetric wrinkled architecture not only imparts distinctive geometric characteristics to the material, but also enhances interfacial phonon scattering, effectively suppressing the lattice thermal conductivity (\kl). To define the transport directions and standardize the simulation domain, a standard unit cell containing eight atoms, as highlighted by the red rectangle in Figure~\ref{fig:structure-elf-cohp}A, is chosen for all analyses in this work.

\begin{table}[htp]
\caption{
Lattice constants, lattice angles, bond lengths ($d$), bond angles ($\theta$), and effective thickness ($H_\mathrm{eff}$) of the PbSe/PbTe monolayer heterostructure.
}
\begin{ruledtabular}
\begin{tabular}{lcccccccc}
Material & $a = b$ (\AA) & $\alpha = \beta$ ($^\circ$) & $\gamma$ ($^\circ$) 
         & $d_\mathrm{Pb-Se}$ (\AA) & $d_\mathrm{Pb-Te}$ (\AA) 
         & $\theta_1$ ($^\circ$) & $\theta_2$ ($^\circ$) & $H_\mathrm{eff}$ (\AA) \\
\hline
PbSe/PbTe & 4.300 & 90 & 120 & 2.854 & 3.006 & 97.74 & 91.32 & 6.038 \\
\end{tabular}
\end{ruledtabular}
\label{tab:lattice_params}
\end{table}

Bonding characteristics play a crucial role in determining the performance of thermoelectric materials. To gain deeper insight into the bonding behavior in the \ce{PbSe/PbTe} monolayer heterostructure and its impact on lattice thermal conductivity ($\kappa_{L}$), we further computed the electron localization function (ELF), Bader charge and crystal orbital Hamilton population (COHP)~\cite{ref32,ref41}. The ELF values provide a measure of electron localization: ELF = 1 corresponds to perfect localization (e.g., lone pair electrons or covalent bonds), ELF = 0.5 represents a free electron gas, and ELF = 0 indicates complete delocalization. As shown in Figure 1B, the ELF values for both \ce{Pb-Se} and \ce{Pb-Te} bonds are relatively low, indicating a weak degree of electron localization. This suggests that the bonding nature is not purely covalent but contains weakly ionic characteristics.To further analyze the bonding characteristics, the Bader charges of the atoms were calculated. The Bader (static) charges calculated for Te, Pb, and Se atoms are \SI{-0.43}{e}, \SI{+0.44}{e}/\SI{+0.63}{e}, and \SI{-0.64}{e}, respectively. These values are significantly smaller in magnitude compared to their nominal oxidation states: \ch{Te} and \ch{Se} typically carry a charge of \SI{-2}{e}, while \ch{Pb} commonly exhibits \SI{+2}{e} or \SI{+4}{e} oxidation states. Specifically, the charge on \ch{Te} is \SI{-0.43}{e}, which corresponds to approximately 21.5\% of its nominal ionic charge (\SI{-2}{e}), indicating that about 78.5\% of the expected ionic charge is not transferred. Similarly, the Bader charge on \ch{Se} is \SI{-0.64}{e}, representing roughly 32\% of the ideal ionic charge, with about 68\% of the charge remaining shared or delocalized. For \ch{Pb}, due to its presence at different atomic positions, the average Bader charge is \SI{+0.535}{e}. These values are significantly lower than the nominal \SI{+2}{e} or \SI{+4}{e} oxidation states, accounting for only about 26.8\% of the ideal ionic charge.This indicates that \ch{Pb} atoms retain the majority of their valence electrons, with Bader charges close to neutrality, reflecting strong electron delocalization. The partial negative charge accumulation on \ch{Te} and \ch{Se}, significantly less than full ionic charges, further supports the presence of substantial covalent bonding character. These observations are consistent with the ELF analysis results. Overall, the combination of partial electron transfer and relatively low ELF values suggests that the \ch{Pb-Se} and \ch{Pb-Te} bonds are polar covalent, characterized by partial electron sharing alongside minor ionic contributions.Furthermore, COHP projection analysis (Figure~\ref{fig:structure-elf-cohp}C) reveals significantly negative -COHP values for \ch{Pb-Se} and \ch{Pb-Te} bonds in the energy range of about \SIrange{-2}{-3}{\electronvolt} below the Fermi level, indicating the presence of dominant antibonding states. These antibonding states weaken the bond strength and lead to pronounced repulsive interactions. The reduction in bond strength and uneven distribution of bond energies disrupt lattice periodicity and phonon propagation, enhancing phonon scattering and further reducing $\kappa_{L}$.In summary, the \ch{Pb-Se} and \ch{Pb-Te} bonds in the \ch{PbSe/PbTe} monolayer heterostructure are mainly polar covalent with weak ionic character. Partial charge transfer, low ELF values, and the occupation of antibonding states collectively contribute to weakened bond strength and enhanced lattice anharmonicity, which are key electronic structure origins for achieving low lattice thermal conductivity in this system.

\begin{figure}
  \centering
  \includegraphics[angle=0,width=0.7\textwidth]{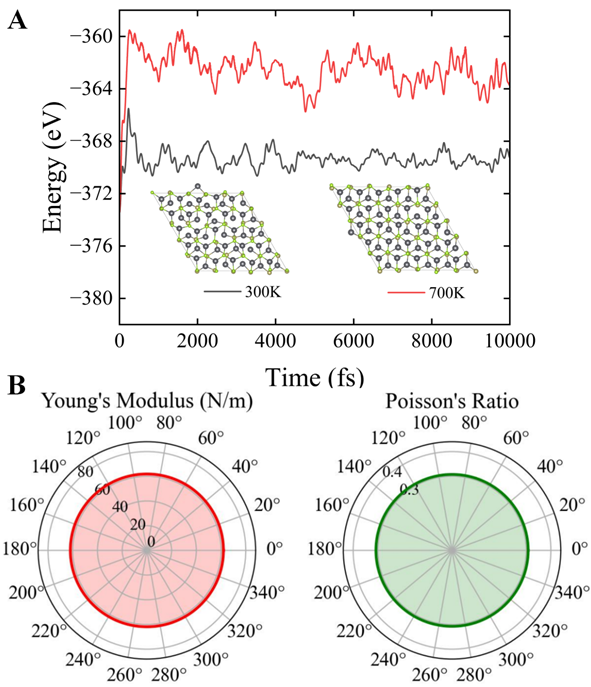}
  \caption{\label{fig:aimd-yang}  
    (A) Ab initio molecular dynamics (AIMD) results of the 4 $\times$ 4 $\times$ 1 supercell of PbSe/PbTe monolayer heterostructure at 300 K and 700 K.  
    (B) Young’s modulus and Poisson’s ratio of the PbSe/PbTe monolayer heterostructure.  
  }
\end{figure}

Good stability is essential for thermoelectric materials. Firstly, the elastic constants matrix of the \ce{PbSe/PbTe} monolayer heterostructure was calculated, as shown in Table~\ref{tab:elastic}. The \ce{PbSe/PbTe} monolayer heterostructure satisfies the Born-Huang criteria for mechanical stability, namely:$C_{11} > 0$; $C_{66}> 0$; $C_{11}C_{22} - C_{12}^2 > 0$. Therefore, the \ce{PbSe/PbTe} monolayer is mechanically stable. Furthermore, the relationship among Young's modulus ($Y$), Poisson's ratio ($\nu$), and elastic constants is given by~\cite{ref42}:

\begin{align}
Y(\theta) &= \frac{C_{11}C_{22} - C_{12}^2}{C_{11} \sin^4\theta + A \sin^2\theta \cos^2\theta + C_{22} \cos^4\theta} \tag{6} \\
\nu(\theta) &= \frac{C_{12} \sin^4\theta - B \sin^2\theta \cos^2\theta + C_{12} \cos^4\theta}{C_{11} \sin^4\theta + A \sin^2\theta \cos^2\theta + C_{22} \cos^4\theta} \tag{7}
\end{align}

where $A = (C_{11}C_{22} - C_{12}^2)/C_{66} - 2C_{12}$, $B = C_{11} + C_{22} - (C_{11}C_{22} - C_{12}^2)/C_{66}$, $C_{66} = (C_{11} - C_{12})/2$, and $\theta$ denotes the polar angle measured from the $x$-axis. Based on the polar coordinate representation shown in Figure~\ref{fig:aimd-yang}B, the Young's modulus of the \ce{PbSe/PbTe} monolayer heterostructure is \SI{61.94}{\newton\per\meter}, significantly lower than common 2D materials such as h-\ce{BN} (\SI{270}{\newton\per\meter}) and graphene ($350 \pm 3.15\,\si{\newton\per\meter}$)~\cite{ref43,ref44}. The Poisson's ratio of the \ce{PbSe/PbTe} heterostructure is 0.307, higher than that of h-\ce{BN} (0.211) and graphene (0.175)~\cite{ref43,ref44}. This indicates that the \ce{PbSe/PbTe} monolayer exhibits lower stiffness and greater flexibility compared to other common two-dimensional materials. Additionally, the Debye temperature of the \ce{PbSe/PbTe} monolayer was calculated to be \SI{43.9}{\kelvin}. Compared with the Debye temperatures of black phosphorus (\SI{500}{\kelvin})~\cite{ref45} and h-\ce{MoS2} (\SI{600}{\kelvin})~\cite{ref46}, the lower Debye temperature of \ce{PbSe/PbTe} suggests a correspondingly low lattice thermal conductivity.

\begin{table}[htp]
\caption{
Elastic stiffness matrix ($C$), 2D Young's modulus ($Y^{2D}$), Poisson's ratio ($\nu$), and Debye temperature ($\Theta$) of the PbSe/PbTe monolayer heterostructure.
}
\begin{ruledtabular}
\begin{tabular}{lccccccc}
Material & $C_{11}$ (GPa) & $C_{12}$ (GPa) & $C_{22}$ (GPa) & $C_{66}$ (GPa) 
         & $Y^{2D}$ (N/m) & $\nu$ & $\Theta$ (K) \\
\hline
PbSe/PbTe & 9.208 & 2.849 & 9.202 & 3.174 & 61.94 & 0.307 & 43.9 \\
\end{tabular}
\end{ruledtabular}
\label{tab:elastic}
\end{table}

To further assess the thermal stability of the PbSe/PbTe monolayer heterostructure, ab initio molecular dynamics (AIMD) simulations were performed in the canonical (NVT) ensemble using a \( 4 \times 4 \times 1 \) supercell at 300 K and 700 K, with a total simulation time of 10 ps and a time step of 1 fs. The results are presented in Figure~\ref{fig:aimd-yang}A, indicating that the total energy remained relatively constant throughout the simulations, and the supercell crystal structure exhibited no distortion or disintegration. These findings confirm the structural stability of the PbSe/PbTe monolayer heterostructure at both room temperature (300 K) and elevated temperature (700 K).

\subsection{Phonon Transport Properties}

\begin{figure}
  \centering
  \includegraphics[angle=0,width=0.8\textwidth]{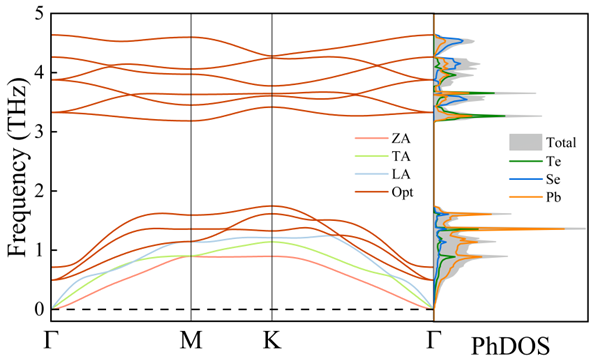}
  \caption{\label{fig:3}  
    Phonon dispersion curves and PhDOS of the PbSe/PbTe monolayer heterostructure. “Opt” represent the optical phonon modes.}
\end{figure}

Based on the machine learning MTP method, we systematically evaluated the phonon dispersion characteristics of the PbSe/PbTe monolayer heterostructure. This method achieves a favorable balance between computational accuracy and efficiency, and its effectiveness in thermoelectric materials research has been extensively validated by numerous studies. Moreover, our previous work has confirmed its applicability to similar systems~\cite{ref47,ref48}. Furthermore, as shown in Supplementary Figure 4, the phonon dispersion curves of the PbSe/PbTe monolayer heterostructure unit cell calculated using the finite displacement method and the MTP potential exhibit excellent agreement, which fully validates the reliability of the MTP approach employed in predicting the phonon transport properties of the PbSe/PbTe monolayer heterostructure. Figure~\ref{fig:3} shows, the phonon dispersion relations and corresponding phonon density of states (PhDOS) of the PbSe/PbTe monolayer heterostructure along the high-symmetry path $\Gamma$-M-K-$\Gamma$. The system comprises three acoustic phonon branches, corresponding to the out-of-plane (ZA), transverse (TA), and longitudinal (LA) vibrational modes, while the remaining nine branches represent optical modes. No imaginary frequencies appear throughout the Brillouin zone, indicating dynamical stability of the structure. Further analysis of the phonon spectrum reveals that the highest phonon frequency of the material is 4.72 THz, significantly lower than typical thermoelectric materials such as ZnSe ($\sim$8 THz) and MoS2 ($\sim$14 THz)~\cite{ref49,ref50}. This low-frequency characteristic typically indicates a reduced lattice thermal conductivity (\kl). The acoustic branches are concentrated within the 0$\sim$1.2 THz range, whereas the optical branches mainly lie between 1.2 and 4.7 THz, exhibiting a distinct acoustic–optical phonon gap. Notably, the acoustic modes in the phonon dispersion curves show pronounced flatness, especially the ZA branch, whose dispersion is the flattest, implying an extremely low phonon group velocity. It is well known that lower phonon group velocities suppress lattice thermal conductivity. Therefore, the synergistic effects of frequency range, mode coupling, and phonon transport characteristics in the PbSe/PbTe monolayer heterostructure collectively contribute to the effective reduction of \kl.

A peak in the phonon density of states around ~1.3 THz is observed, indicating strong acoustic–optical phonon coupling in the PbSe/PbTe monolayer heterostructure. The intense coupling between acoustic and optical modes induces strong phonon scattering, thereby reducing \kl. In addition, Pb atoms exhibit dominant vibrational contributions in the low-frequency acoustic region ($\sim$0 to 1.7 THz), and their contributions in the mid- to high-frequency optical region ($\sim$3.2 to 4.8 THz) are also non-negligible. This phenomenon may originate from the complex vibrational coupling induced by weak Pb–-Pb bonds between the top and bottom Pb layers, which enhances both phonon–phonon and acoustic–optical phonon scattering mechanisms, further suppressing \kl.

\begin{figure}
  \centering
  \includegraphics[angle=0,width=0.7\textwidth]{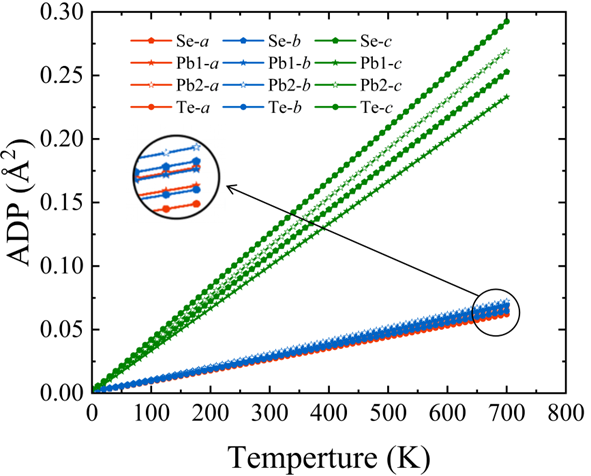}
  \caption{\label{fig:4}  
    ADP of different atoms along various directions in the PbSe/PbTe monolayer heterostructure.}
\end{figure}

Intense atomic thermal vibrations are often closely related to strong anharmonicity in the crystal structure. To further quantify the degree of anharmonicity and characterize atomic vibrations in the PbSe/PbTe monolayer heterostructure, we computed the anisotropic displacement parameters (ADP) of atoms along different crystallographic axes, as shown in Figure~\ref{fig:4}. It is clearly observed that the ADP along the $c$-axis are significantly larger than those along the $a$- and $b$-axes, indicating more intense thermal vibrations perpendicular to the plane. Specifically, Pb atoms in the top and bottom layers exhibit relatively smaller ADP than Te atoms, which is mainly attributed to their larger atomic mass~\cite{ref51}, whereas the higher ADP values of Te atoms can be ascribed to the weaker bonding strength with surrounding atoms. This finding is consistent with the previous conclusions from ELF and –COHP analyses, indicating weak chemical bonding interactions between Pb and Te, which allows freer atomic thermal motion and thereby enhances the structural anharmonicity. 

Overall, the pronounced atomic vibration amplitudes along the $c$-axis, the local bonding strength heterogeneity, and the large ADP values in the PbSe/PbTe monolayer heterostructure collectively indicate strong anharmonicity of the system, which effectively enhances phonon scattering and consequently significantly reduces the lattice thermal conductivity (\kl).

From the previous analyses, it is evident that the PbSe/PbTe monolayer heterostructure exhibits considerable anharmonicity, and we therefore predict that the PbSe/PbTe monolayer heterostructure will possess a low lattice thermal conductivity (\kl). To more accurately evaluate the thermal transport properties of the PbSe/PbTe monolayer heterostructure, we introduced a four-phonon scattering model in addition to the three-phonon scattering processes, thereby enabling a comprehensive assessment of phonon-mediated thermal transport. Based on the phonon Boltzmann transport equation (BTE), the lattice thermal conductivity can be expressed as:

\begin{equation}
K_{L,k} = \frac{1}{V}\sum_{k} C_{kq} \tau_{kq} v_{kq}^2
\end{equation}

\begin{figure}
  \centering
  \includegraphics[angle=0,width=0.7\textwidth]{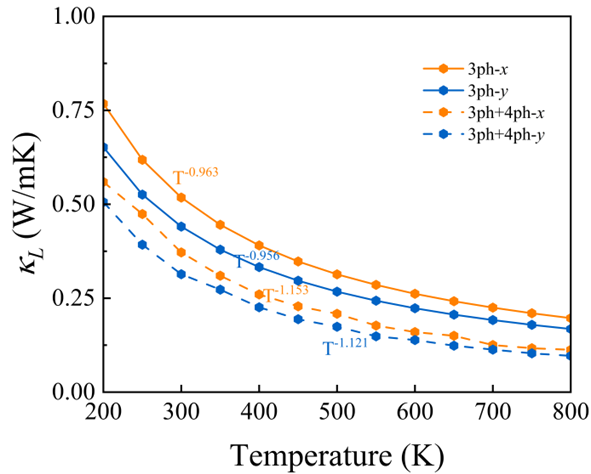}
  \caption{\label{fig:5}  
     Temperature dependence of lattice thermal conductivity (\kl) of the PbSe/PbTe monolayer heterostructure along the $x$ and $y$ directions considering three-phonon and four-phonon scattering mechanisms.}
\end{figure}

Here, $C_{kq}$, $v_{kq}$ and $\tau_{kq}$ denote the mode-specific heat capacity, phonon group velocity, and phonon relaxation time of the $\bm{k}$-th phonon branch, respectively. As shown in Figure~\ref{fig:5}, the temperature dependence of the lattice thermal conductivity (\kl) of the monolayer PbSe/PbTe monolayer heterostructure was systematically analyzed under different scattering mechanisms, where the solid lines represent results considering only three-phonon scattering processes, and the dashed lines include the four-phonon scattering mechanism as well. Overall, \kl  gradually decreases with increasing temperature, which is primarily attributed to the enhanced phonon-phonon scattering strength at elevated temperatures. Due to the pronounced anisotropy of the heterostructure, the lattice thermal conductivity along the $y$-direction remains consistently lower than that along the $x$-direction, reflecting the directional dependence of its lattice dynamical properties. At 300 K, When the three- and four-phonon scattering processes are considered, the thermal conductivities of the lattice $\kappa_{\rm L}$ along the $x$ and $y$ directions are as low as \SI{0.37}{\watt\per\meter\per\kelvin} and \SI{0.31}{\watt\per\meter\per\kelvin}, respectively, representing decreases of approximately 29\% and 10\% compared to \SI{0.52}{\watt\per\meter\per\kelvin} and \SI{0.44}{\watt\per\meter\per\kelvin} obtained with only three-phonon scattering. This significant reduction is attributed to the introduction of additional energy dissipation channels by four-phonon scattering processes, which further enhance phonon scattering intensity and effectively suppress the increase in thermal conductivity.To further understand the influence of four-phonon scattering, we fitted the temperature dependence of $\kappa_{\rm L}$ using a power-law relationship ($\kappa_{\rm L} \sim T^{-\alpha}$), as shown in Figure~5. In the three-phonon-only case, the extracted exponents are $\alpha_{x} = 0.963$ and $\alpha_{y} = 0.956$, which are close to the well-established $T^{-1}$ scaling predicted by the conventional phonon gas model~\cite{ref52}. This agreement indicates that under three-phonon scattering, the system exhibits relatively weak anharmonicity and maintains well-defined phonon quasiparticles.However, upon including four-phonon scattering, the temperature dependence becomes noticeably stronger, with $\alpha_{x} = 1.153$ and $\alpha_{y} = 1.121$. This enhancement reflects the increased phonon-phonon scattering rates at elevated temperatures due to additional anharmonic channels. While the behavior remains consistent with a phonon-mediated transport picture, the stronger temperature sensitivity highlights the importance of higher-order anharmonic interactions.These results underscore the need to include four-phonon processes for accurate thermal conductivity predictions, especially under high-temperature conditions. Therefore, four-phonon scattering plays a crucial role in the thermal transport behavior of the material studied in this work and must be thoroughly considered when accurately evaluating the $\kappa_{\rm L}$.

\begin{figure}
  \centering
  \includegraphics[angle=0,width=0.7\textwidth]{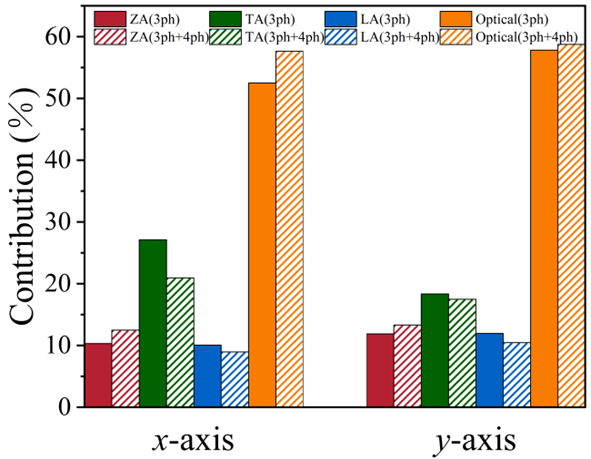}
  \caption{\label{fig:6}  
    Contributions of different phonon modes (ZA, TA, LA, and Optical) to \kl of the PbSe/PbTe monolayer heterostructure, including four-phonon scattering in both the $x$ and $y$ directions. Solid-filled bars represent results obtained using the three-phonon scattering model, while hatched-filled bars denote results considering four-phonon scattering.}
\end{figure}

To further elucidate the phonon transport mechanisms in the monolayer PbSe/PbTe monolayer heterostructure, we performed a detailed analysis of the contributions of different acoustic modes (ZA, TA, LA) and optical modes to the total lattice thermal conductivity ($\kappa_{\textrm{L}}$). As shown in Figure~\ref{fig:6}, at \SI{300}{\kelvin}, the contributions of each phonon branch to $\kappa_{\textrm{L}}$ were evaluated under both three-phonon scattering only, and the combination of three- and four-phonon scattering models. The results indicate that regardless of the phonon scattering model employed, the contribution of optical branches to the $\kappa_{\textrm{L}}$ consistently exceeds \SI{50}{\percent}. Generally, acoustic phonon modes are the primary contributors to $\kappa_{\textrm{L}}$~\cite{ref53}, however, it is noteworthy that in the PbSe/PbTe monolayer heterostructure, the optical modes play a dominant role. Specifically, under the four-phonon scattering model, the contribution of optical modes along the $y$-direction can reach up to \SI{58}{\percent}. In the PbSe/PbTe monolayer heterostructure, however, the optical phonons play a dominant role, which can be attributed to several key factors. As shown in Figure~\ref{fig:10}A, some optical modes exhibit exceptionally high group velocities exceeding \SI{1.5}{\kilo\meter\per\second}, even surpassing those of the ZA and TA acoustic modes. This challenges the traditional assumption that optical phonons contribute little due to low group velocity. Additionally, Figure~\ref{fig:10}B reveals that the Gr\"uneisen parameters of optical phonons are moderate and smoothly distributed, indicating limited anharmonicity and reduced scattering strength. Despite their relatively higher scattering rates (Supplementary Figure 5), the combination of high mode density, large group velocity, and manageable anharmonicity enables optical phonons to become the principal heat carriers in this system.

\begin{figure}
  \centering
  \includegraphics[angle=0,width=0.7\textwidth]{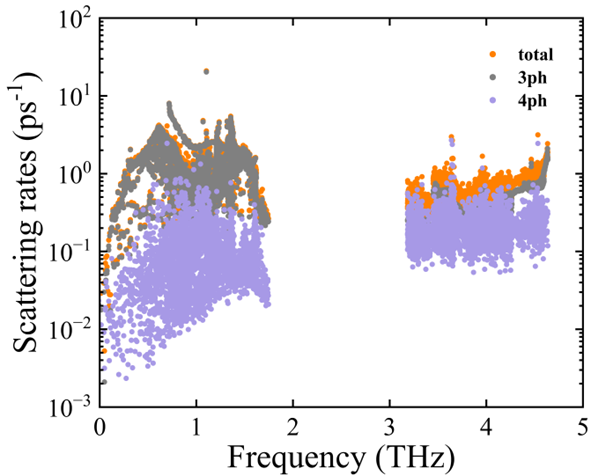}
  \caption{\label{fig:7}  
    Frequency-dependent three-phonon and four-phonon scattering rates of PbSe/PbTe monolayer heterostructure.}
\end{figure}

To verify the validity of the phonon quasiparticle picture, we plotted the total phonon scattering rates from 400 K to 800 K under the four-phonon scattering model, and included a reference red line corresponding to the breakdown threshold of the quasiparticle picture, defined as $1/\tau_{\rm ph} = \omega_{\rm ph}/2\pi$. When the total scattering rate exceeds this curve, the phonon lifetime is shorter than its vibrational period, and the quasiparticle description fails~\cite{ref54}. As shown in Supplementary Figure 5, although a few low-frequency modes exceed this threshold at higher temperatures, the majority of phonon modes remain within an acceptable range, confirming the validity of using the Boltzmann transport equation in this study. Next, to gain deeper insight into the low thermal conductivity exhibited by the PbSe/PbTe monolayer heterostructure, we further analyzed its anharmonic phonon-phonon scattering behavior. As shown in Figure~\ref{fig:7}, the frequency-dependent three-phonon and four-phonon scattering rates of the PbSe/PbTe heterostructure at \SI{300}{\kelvin} are presented. Overall, the anharmonic scattering rates of this material span the range from $10^{-2}$ to $10^{1}$~\si{\per\pico\second}. Compared to some reported highly anharmonic thermoelectric materials, the scattering rates are significantly higher than those of materials with relatively low anharmonic scattering levels like GeS~\cite{ref55}, KBaBi~\cite{ref56} and Na$_2$TiSb~\cite{ref57} (which typically exhibit rates below \SI{1}{\per\pico\second}), indicating the pronounced anharmonicity of the PbSe/PbTe heterostructure.

\begin{figure}
  \centering
  \includegraphics[angle=0,width=0.99\textwidth]{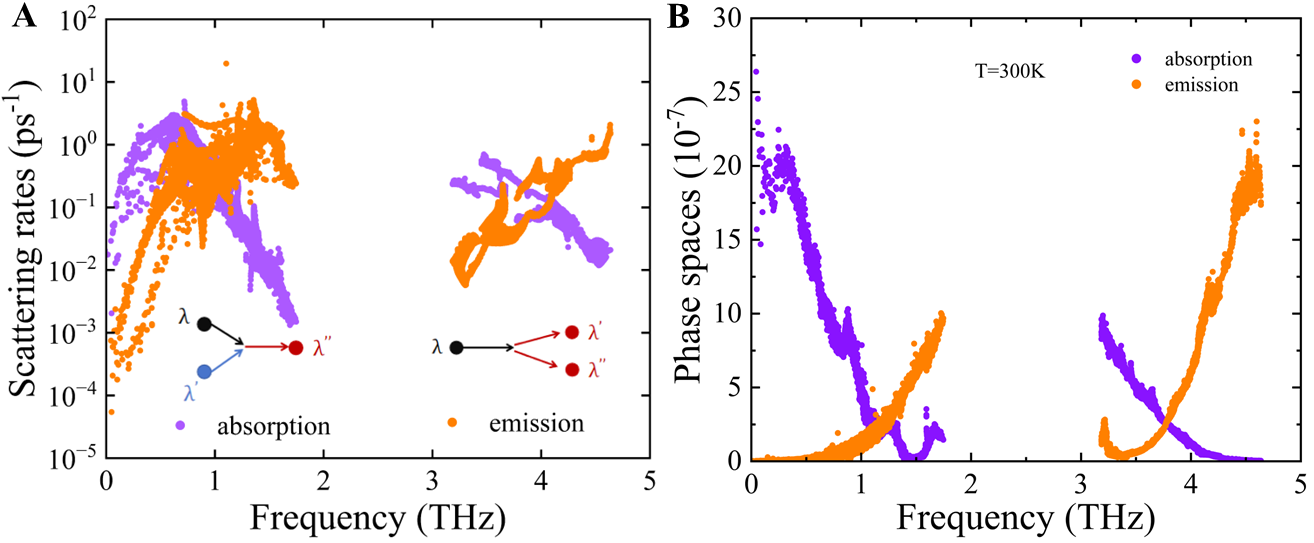}
  \caption{\label{fig:8}  
    Absorption and emission processes under three-phonon scattering of PbSe/PbTe monolayer heterostructure: (A) scattering rates; (B) phase space as a function of frequency.}
\end{figure}

As the frequency increases, both three-phonon and four-phonon scattering rates exhibit a gradually increasing trend, reaching peaks in certain frequency ranges before stabilizing, with two distinct scattering peaks observed near approximately \SI{0.5}{\tera\hertz} and \SI{1.5}{\tera\hertz}, respectively. Based on phonon density of states analysis, as shown in Figure~\ref{fig:3}, the scattering peak around \SI{0.5}{\tera\hertz} is primarily attributed to the low-frequency vibrational modes of Pb atoms, where strong coupling occurs between the lowest optical branch and acoustic branches, thereby facilitating enhanced phonon-phonon scattering processes. Although the magnitudes of three- and four-phonon scattering rates are comparable, the latter exhibits significant influence across multiple frequency ranges, indicating that four-phonon scattering is also a critical mechanism governing lattice thermal conductivity in the anharmonic thermal transport of the PbSe/PbTe monolayer heterostructure.
Figure~\ref{fig:8}A illustrates the scattering channel behavior under three-phonon scattering mechanisms, distinguishing between absorption processes ($\lambda+\lambda^{\prime}\rightarrow\lambda^{\prime\prime}$) and emission processes ($\lambda\rightarrow\lambda^{\prime}+\lambda^{\prime\prime}$). With increasing frequency, the scattering rate of the emission process significantly increases, whereas that of the absorption process exhibits a decreasing trend. This behavior indicates that the large optical phonon gap effectively suppresses the activation of absorption-type three-phonon scattering channels, thereby revealing the complexity of dynamic equilibrium in phonon interactions and energy transfer~\cite{ref58}.

To further understand the frequency dependence of these scattering channels, the three-phonon scattering phase space was computed and is presented in Figure~\ref{fig:8}B, separately for absorption and emission processes. It is evident that in the low-frequency range (\SIrange{0}{1}{\tera\hertz}), the absorption process possesses a broad scattering phase space, while a relatively large emission phase space exists in the high-frequency range (approximately \SIrange{4}{4.8}{\tera\hertz}). As shown in Figure~\ref{fig:3}, phonon density of states analysis indicates that the low-frequency region is mainly contributed by acoustic modes of Pb atoms, whereas the high-frequency region is dominated by optical vibrations primarily involving Pb and Se atoms. These results suggest that the optical branches involving Pb and Se atoms provide abundant three-phonon scattering channels at high frequencies, which enhances the density of scattering events and plays a critical role in reducing the thermal conductivity.

\begin{figure}
  \centering
  \includegraphics[angle=0,width=0.7\textwidth]{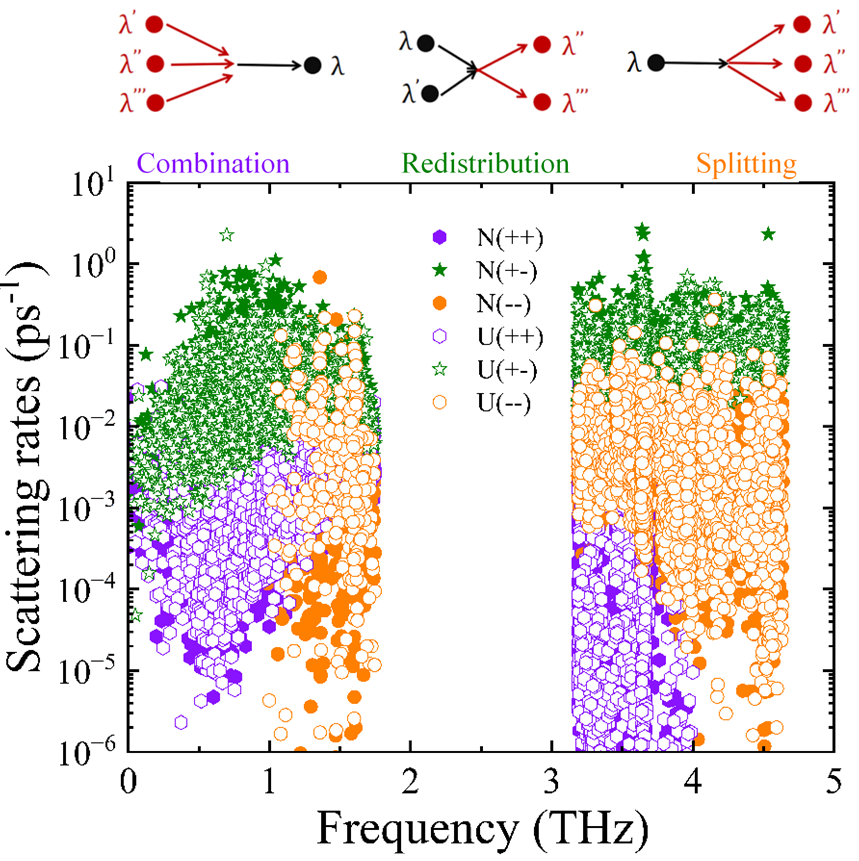}
  \caption{\label{fig:9}  
    Calculated four-phonon scattering rates of PbSe/PbTe heterostructure monolayer as a function of phonon frequency, including three scattering channels: combination (++), redistribution (+-), and splitting (- -)."U" denotes Umklapp processes, while "N" refers to the normal processes.}
\end{figure}

\begin{figure}
  \centering
  \includegraphics[angle=0,width=0.99\textwidth]{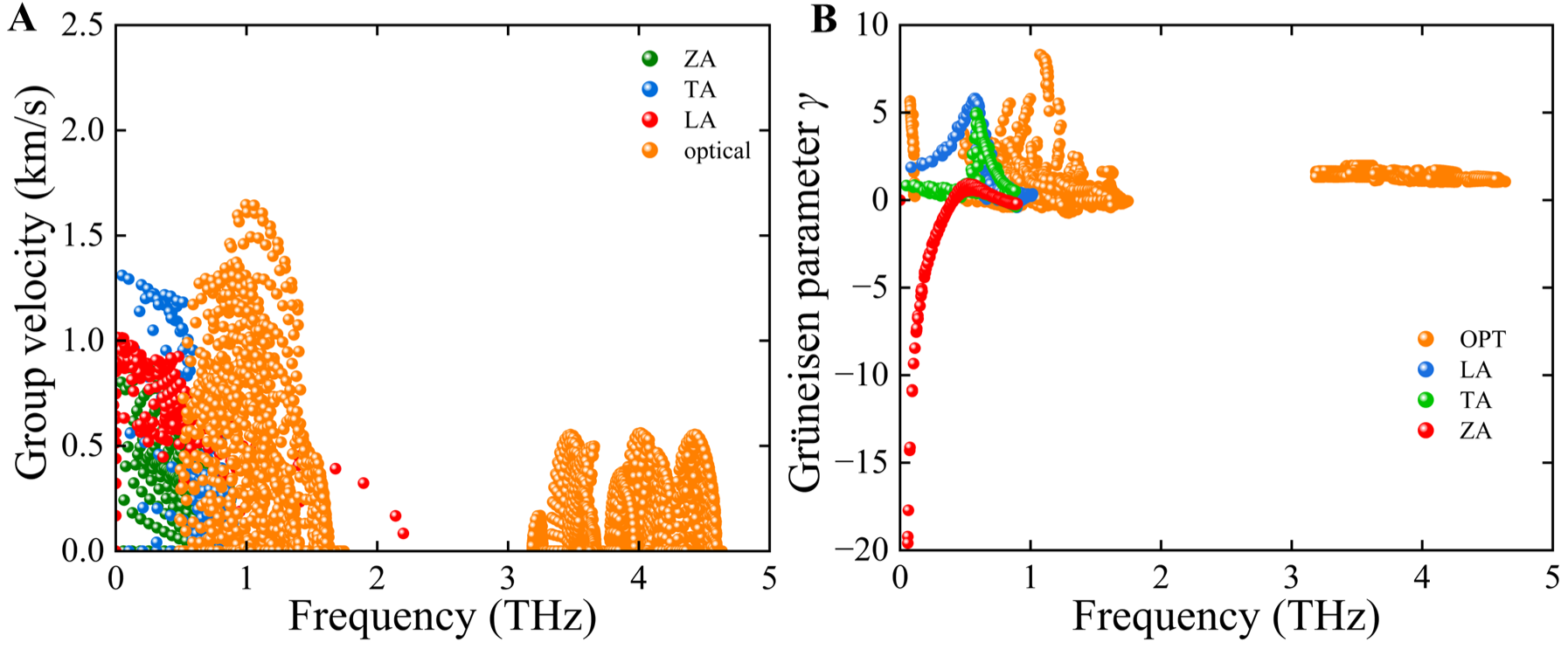}
  \caption{\label{fig:10}  
	(A) Phonon group velocity ($v_g$) and (B) Gr\"uneisen parameter ($\gamma$) as functions of phonon frequency for \ce{PbSe/PbTe} monolayer heterostructure.}
\end{figure}

To gain deeper insight into four-phonon scattering mechanisms, we further investigated all possible scattering channels in the monolayer \ce{PbSe/PbTe} heterostructure. As shown in Figure~\ref{fig:9}, the four-phonon scattering processes in the \ce{PbSe/PbTe} monolayer heterostructure include both Umklapp and Normal types. These processes can be classified into three main scattering channels: combination (++ process, $\lambda'+\lambda''+\lambda'''\rightarrow\lambda$), redistribution (+-- process, $\lambda+\lambda^{\prime}\rightarrow\lambda''+\lambda'''$) and splitting (--- process, $\lambda\rightarrow\lambda^{\prime}+\lambda''+\lambda'''$), where $\lambda$ represents a phonon mode. It can be observed that the Umklapp and Normal processes contribute comparably. Although normal processes do not directly resist heat flow, they influence thermal conductivity via phonon-momentum redistribution. In contrast, Umklapp processes introduce direct thermal resistance and reduce the lattice thermal conductivity. The Umklapp process clearly dominates the four-phonon scattering contribution in the \ce{PbSe/PbTe} heterostructure. More importantly, among these scattering events, the redistribution process plays the most significant role. This dominant channel contributes substantially to the ultralow lattice thermal conductivity observed in the \ce{PbSe/PbTe} monolayer heterostructure~\cite{ref59}.

Moreover, in the \ce{PbSe/PbTe} monolayer heterostructure, the strong four-phonon scattering is intrinsically associated with the presence of an acoustic-optical phonon band gap, as shown in Figure~\ref{fig:3}. In the vicinity of the pronounced gap between the acoustic and optical branches (\SIrange{1.7}{3.2}{\tera\hertz}), the available phase space for three-phonon scattering processes is significantly reduced. As illustrated in Figures~\ref{fig:8}A and~\ref{fig:8}B, this leads to a notable suppression of both the three-phonon phase space and the corresponding scattering rates. Specifically, absorption-type three-phonon processes are largely prohibited due to the strict energy conservation constraints imposed by the gap. In contrast, four-phonon processes particularly redistribution and splitting channels remain active near the gap region, as demonstrated in Figure~\ref{fig:9}. These higher-order interactions are less constrained by energy selection rules and are capable of bridging the phonon band gap, thereby introducing additional anharmonic scattering pathways that are forbidden in the three-phonon regime. This mechanism is further supported by Figure~\ref{fig:7}, which shows that while the contribution of four-phonon processes is generally lower than that of three-phonon processes in the low-frequency region, a noticeable increase in four-phonon scattering occurs near the gap where three-phonon scattering is substantially weakened. This results in a comparable contribution from both scattering mechanisms in the gap region. These findings highlight the dual role of the phonon band gap: it suppresses conventional three-phonon scattering while simultaneously enhancing the importance of four-phonon interactions. Therefore, the increased activity of four-phonon processes plays a critical role in reducing the lattice thermal conductivity of the PbSe/PbTe monolayer system.

We further calculated the phonon group velocity of the \ce{PbSe/PbTe} monolayer heterostructure at \SI{300}{\kelvin} based on the standard group velocity expression~\cite{ref60}:

\begin{equation}
\mathbf{v}_{\mathrm{g}}=\frac{\partial \mathbf{w}_{k}(\boldsymbol{q})}{\partial \boldsymbol{q}}
\label{eq:group_velocity}
\end{equation}

where $\omega_{k}(q)$ and $q$ denote the phonon frequency and wave vector of the $k$-th mode, respectively. The calculated results are illustrated in Figure~\ref{fig:10}A. It is observed that the group velocity of optical modes is significantly larger than that of the acoustic modes. Among the acoustic branches, the ZA mode exhibits the lowest group velocity, which is consistent with our earlier analysis of the phonon dispersion curves. Interestingly, the group velocity of optical phonons near ~0.8 THz is comparable to that of acoustic phonons. This phenomenon can be attributed to the presence of weak interatomic bonding and harmonic mixing between optical and acoustic branches. From the mode-resolved thermal conductivity contribution analysis, it is evident that the lattice thermal conductivity (\kl) of the PbSe/PbTe heterostructure monolayer is predominantly determined by the optical phonons. Therefore, the group velocity of optical modes plays a crucial role in shaping \kl.
Further analysis reveals that the peak group velocity of the \ce{PbSe/PbTe} monolayer heterostructure reaches only \SI{1.64}{\kilo\meter\per\second}, which is significantly lower than that of well-known thermoelectric materials such as \ce{PbTe} ($\sim$\SI{1.80}{\kilo\meter\per\second}) and \ce{As2Ge} ($\sim$\SI{4.5}{\kilo\meter\per\second})~\cite{ref24,ref61}. Such a low group velocity peak is a typical indicator of low lattice thermal conductivity. The Grüneisen parameter ($\gamma$), which quantifies the degree of anharmonicity in a material, can be expressed as~\cite{ref58}:

\begin{equation}
\gamma_{k}(q) = -\frac{V_{0}}{\omega_{k}(q)} \frac{\partial \omega_{k}(q)}{\partial V}
\label{eq:gruneisen}
\end{equation}

Here, $\omega_{k}$ represents the phonon frequency of the $k$-th mode at the equilibrium volume $V_{0}$. A larger absolute value of the Grüneisen parameter $|\gamma|$ indicates stronger anharmonicity, which typically correlates with lower lattice thermal conductivity $\kappa_{L}$. As shown in Figure 10B, the absolute Grüneisen parameter $|\gamma|$ of the \ce{PbSe/PbTe} monolayer heterostructure at \SI{300}{\kelvin} exhibits relatively high values for both acoustic and optical phonon modes, with a maximum reaching up to 20. This clearly indicates a strong anharmonic phonon scattering behavior in the structure, which is consistent with the previous analyses. Moreover, materials with such pronounced anharmonicity necessitate the inclusion of four-phonon scattering mechanisms in thermal transport evaluations, as their contributions to phonon-phonon scattering and the suppression of \kl become non-negligible.

\begin{figure}
  \centering
  \includegraphics[angle=0,width=0.99\textwidth]{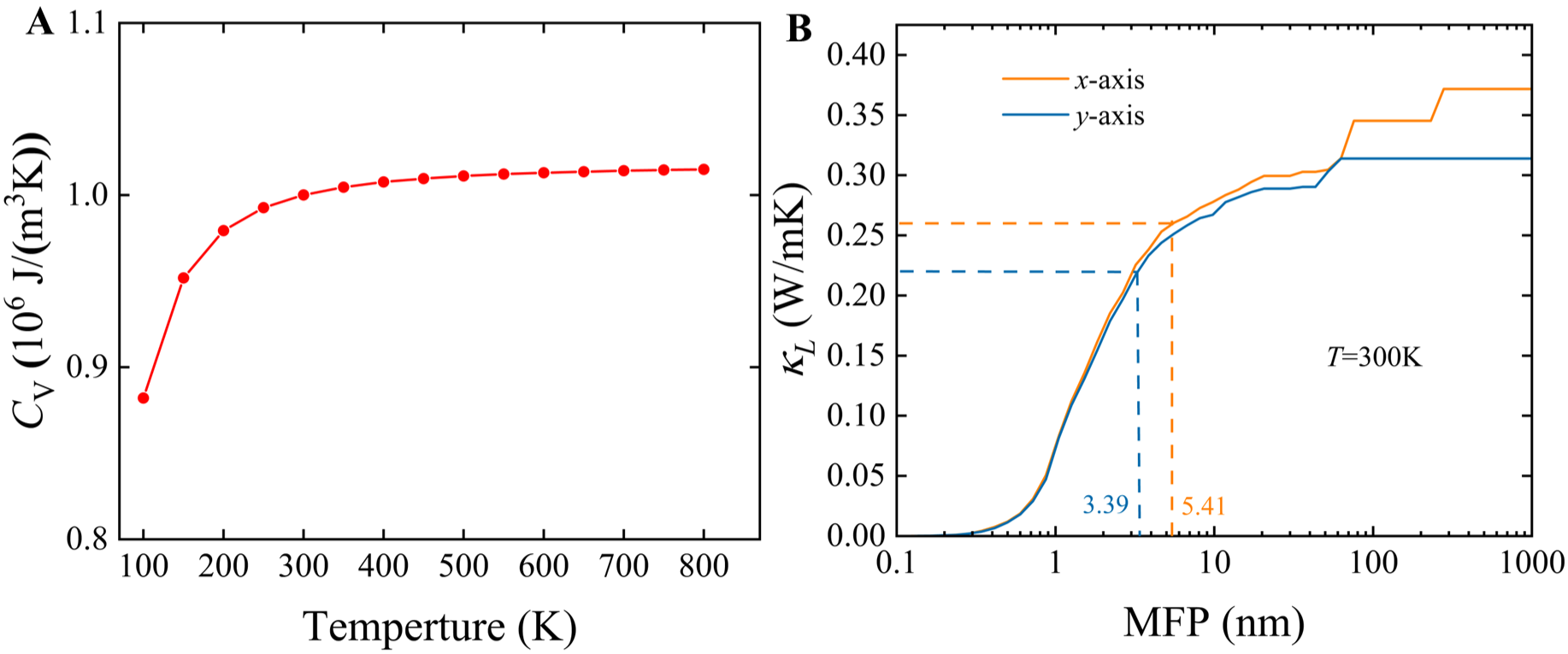}
  \caption{\label{fig:11}  
	(A) Calculated volumetric heat capacity of PbSe/PbTe heterostructure monolayer as a function of temperature; (B) lattice thermal conductivity at 300 K versus phonon mean free path.}
\end{figure}

In addition, we evaluated the specific heat capacity $C_v$, another critical parameter influencing the lattice thermal conductivity $\kappa_{L}$. As shown in Figure~\ref{fig:11}A, $C_v$ of the \ce{PbSe/PbTe} monolayer heterostructure increases gradually with temperature and eventually approaches saturation at high temperatures. Notably, the maximum value of $C_v$ reaches only \SI{1.01e6}{\joule\per\cubic\meter\per\kelvin}, and such a low heat capacity also contributes to the ultralow $\kappa_{L}$ observed in this system.Furthermore, we calculated the phonon mean free path (MFP) of the \ce{PbSe/PbTe} heterostructure monolayer. 

As depicted in Figure~\ref{fig:11}B, the relationship between $\kappa_{L}$ and phonon MFP is illustrated. The critical MFP corresponding to 50\% of the total lattice thermal conductivity is highlighted. Typically, nanostructures exhibit phonon MFPs under \SI{100}{\nano\meter}~\cite{ref60}, however, the values of MFP along the $x$ and $y$ directions for the \ce{PbSe/PbTe} monolayer heterostructure are as small as \SI{5.41}{\nano\meter} and \SI{3.39}{\nano\meter}, respectively. This indicates that the thermal conductivity in this system is relatively insensitive to dimensional scaling. Therefore, conventional structural engineering strategies such as nanostructuring or polycrystallization may have limited effectiveness in reducing $\kappa_{L}$.

\subsection{Electronic Transport Properties}

\begin{figure}
  \centering
  \includegraphics[angle=0,width=0.8\textwidth]{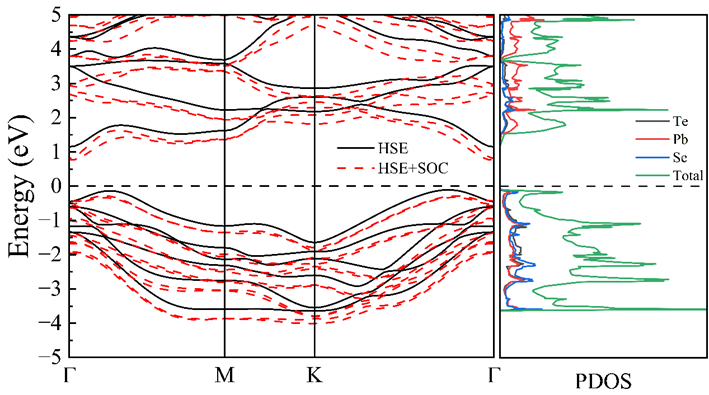}
  \caption{\label{fig:12}  
    Band structures and density of states calculated using HSE and HSE+SOC methods.}
\end{figure}

We next analyzed the electronic transport properties of the \ce{PbSe/PbTe} monolayer heterostructure. Initially, the electronic band structure was calculated using the PBE functional. As shown in Table S1 of the Supporting Information, the \ce{PbSe/PbTe} monolayer heterostructure exhibits a band gap of \SI{0.77}{\electronvolt}, indicating its semiconducting nature. However, it is well known that the PBE functional tends to significantly underestimate band gaps. To obtain more accurate electronic band characteristics, we further performed calculations using the hybrid HSE06 functional. The resulting band structure and density of states (DOS) are presented in Figure~\ref{fig:12}. The \ce{PbSe/PbTe} monolayer heterostructure exhibits an indirect band gap, with the band gap value increasing to \SI{1.25}{\electronvolt}, as listed in Table S1. This increase is primarily attributed to the upward shift of the conduction band upon applying the HSE06 method.

Moreover, the electronic states near the valence band maximum (VBM) are relatively flat, resulting in a large effective mass that contributes to a high Seebeck coefficient. In contrast, the sharp and highly dispersive nature of the conduction band minimum (CBM) favors high carrier mobility. These intrinsic band characteristics are beneficial for achieving excellent thermoelectric (TE) performance. The projected density of states (PDOS) analysis shows that the CBM is primarily contributed by Pb atoms, while the VBM originates mainly from Se atoms. Additionally, the sharp increase in the DOS near the Fermi level is indicative of a potentially enhanced Seebeck coefficient.

Given the heavy atomic mass of Pb in the PbSe/PbTe heterostructure monolayer, we also considered the influence of spin orbit coupling (SOC). Upon including SOC in HSE06 calculations. we observed an overall downward shift in the band structure shown in Figure~\ref{fig:12}. However, the bandgap remains at approximately 1.05 eV, showing no significant deviation from the value without SOC. Therefore, SOC effects can be reasonably neglected in subsequent calculations.

Based on the analysis of the band structure, we predict that the \ce{PbSe/PbTe} monolayer heterostructure possesses promising electronic transport properties. To quantitatively evaluate this, we calculated the Seebeck coefficient ($S$) and electrical conductivity ($\sigma$) of the \ce{PbSe/PbTe} heterojunction in the temperature range of \SIrange{300}{800}{\kelvin} using the semiclassical Boltzmann transport equation as implemented in the BoltzTraP code~\cite{ref40}.

In addition, the rigid band approximation (RBA) was employed to simulate the effect of carrier doping on the electronic performance. The RBA assumes that the overall band structure remains unchanged upon doping, while only the position of the Fermi level shifts accordingly to represent different doping concentrations. These electronic transport properties can be expressed as~\cite{ref40}:
\begin{equation}
\sigma(T,\mu) = e^2 \int_{-\infty}^{+\infty} d\varepsilon \left[ -\frac{\partial f(T,\mu,\varepsilon)}{\partial \varepsilon} \right] \Sigma(\varepsilon)
\label{eq:gruneisen}
\end{equation}
\begin{equation}
S(T,\mu) = \frac{e}{T\sigma(T,\mu)} \int_{-\infty}^{\infty} d\varepsilon \left[ -\frac{\partial f(T,\mu,\varepsilon)}{\partial \varepsilon} \right] \Sigma(\varepsilon) (\varepsilon - \mu)
\label{eq:gruneisen}
\end{equation}

Here, $\mu$ represents the chemical potential, $e$ is the elementary charge of an electron, and $f$ denotes the Fermi-Dirac distribution function of charge carriers. The function $\Sigma(\epsilon)$ refers to the transport distribution function, which is defined as:

\begin{equation}
\Sigma(\epsilon) = \frac{1}{\Omega N_k} \sum_{nk} \tau_{nk}^e |{v}_{nk}|^2 \delta(\epsilon - \epsilon_{nk})
\label{eq:transport_dist}
\end{equation}

Here, ${v}$ represents the electron group velocity. The temperature-dependent relaxation time $\tau$ is determined based on the deformation potential theory. Considering the dominant role of acoustic phonon scattering, the 2D carrier mobility $\mu_{\text{2D}}$ and the electron relaxation time $\tau$ can be described by the following equations~\cite{ref62}:

\begin{equation}
\mu_{\text{2D}} = \frac{2 e h^3 C^{\text{2D}}}{3 k_B T m^* \sqrt{m_i^* m_j^*} E_i^2}
\label{eq:mobility}
\end{equation}

\begin{equation}
\tau = \frac{\mu m^*}{e}
\label{eq:relaxation}
\end{equation}

Here, $C^{\text{2D}}$ denotes the elastic modulus of the two-dimensional material, defined as:

\begin{equation}
C^{\text{2D}} = \left[ \frac{\partial^2 E}{\partial \rho^2} \right]/S
\label{eq:elastic_modulus}
\end{equation}

where $\rho$ is the uniaxial strain applied along the corresponding direction, and $S$ is the cross-sectional area projected along the $z$-axis. The effective mass $m^*$ along the transport direction is defined as:
The effective mass $m^*$ along the transport direction is defined as:

\begin{equation}
m^* = \frac{\hbar^2}{\partial^2 \varepsilon / \partial k^2}
\label{eq:effective_mass}
\end{equation}

Here, $\varepsilon$ denotes the energy of the electronic band, $\hbar$ is the reduced Planck constant, and $k$ represents the electron wavevector. The deformation potential constant $E$ is defined as 
$E = dE_e/d\rho$, where $E_e$ denotes the energy at the band edge, corresponding to the valence band maximum (VBM) and conduction band minimum (CBM). The elementary charge is denoted as $e$, and the geometric mean of the effective masses along the $x$- and $y$-directions is given as $\sqrt{m_x^* m_y^*}$. $k_B$ stands for the Boltzmann constant. It should be noted that in this work, the carrier relaxation time is calculated using the deformation potential (DP) theory, which primarily considers phonon scattering while neglecting other electron scattering mechanisms, such as optical phonon and impurity scattering. While this simplification may lead to a moderate overestimation of the relaxation time, the DP method has been widely adopted in previous studies due to its computational efficiency and physical relevance, especially for two-dimensional (2D) materials. Given the high computational cost of first-principles electron-phonon coupling calculations, the DP approximation offers a practical balance between accuracy and efficiency. Moreover, numerous prior works have successfully employed the DP theory to study charge transport and thermoelectric properties in 2D systems~\cite{ref42,ref47,ref60}.Therefore despite its limitations, the DP model remains a valuable tool for the preliminary evaluation of carrier mobility and thermoelectric performance in 2D materials~\cite{ref63}. As summarized in Table~\ref{tab:transport_props}, a significant difference in carrier mobility is observed between electrons and holes. This trend is consistent with the band structure analysis, which indicates that the effective mass of holes is larger than that of electrons. Notably, the high electron mobility in the PbSe/PbTe heterostructure monolayer is comparable to that of MoS2. Furthermore, at 300 K, the relaxation time of electrons is noticeably longer than that of holes, leading to superior electronic transport properties under n-type doping. Both the high electron mobility and the relatively long relaxation time are beneficial for enhanced electrical transport performance.

\begin{table}[htp]
\caption{
Calculated effective masses ($m^*$), 2D elastic modulus ($C^{2D}$), deformation potential constants ($E$), carrier mobilities ($\mu_{2D}$), and relaxation times ($\tau$) of the PbSe/PbTe monolayer heterostructure.
}
\begin{ruledtabular}
\begin{tabular}{lccccccc}
Material & Direction & Carrier Type & $C^{2D}$ (J/m$^2$) & $E$ (eV) & $m^{*}$ ($m_0$) & $\mu_{2D}$ (cm$^2$/V$\cdot$s) & $\tau$ ($10^{-14}$s) \\
\hline
PbSe/PbTe & $x$ & n-type & 34.37 & 6.620 & 0.122 & 1126.8 & 7.81 \\
          & $x$ & p-type & 34.37 & 4.134 & 0.627 & 118.87 & 4.23 \\
          & $y$ & n-type & 37.87 & 6.616 & 0.122 & 1243.0 & 8.61 \\
          & $y$ & p-type & 37.87 & 3.312 & 0.531 & 240.95 & 7.27 \\
\end{tabular}
\end{ruledtabular}
\label{tab:transport_props}
\end{table}

\begin{figure}
  \centering
  \includegraphics[angle=0,width=0.88\textwidth]{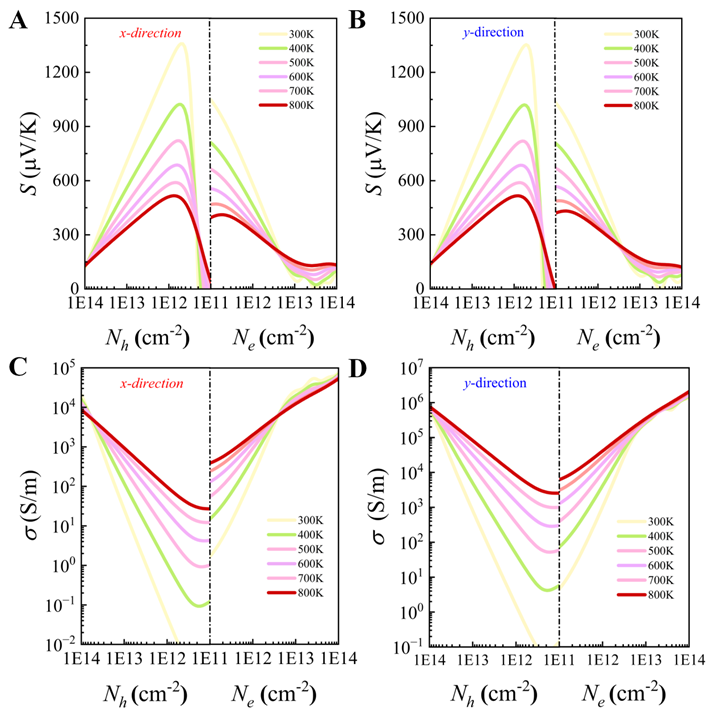}
  \caption{\label{fig:13}  
     Seebeck coefficient ($S$) versus carrier concentration of PbSe/PbTe monolayer heterostructure along $x$ and $y$ directions in the temperature range 300 K to 800 K: (A) p-type; (B) n-type. Electrical conductivity ($\sigma$) versus carrier concentration: (C) p-type; (D) n-type.}
\end{figure}

The Seebeck coefficients ($S$) of p-type and n-type \ce{PbSe/PbTe} monolayer heterostructure are shown in Figure~\ref{fig:13}A and~\ref{fig:13}B. Clearly, $S$ increases with rising temperature, and the $S$ values for n-type doping are significantly lower than those for p-type doping, which is consistent with our previous analysis of the effective masses at the VBM and CBM.At \SI{300}{\kelvin}, the maximum $S$ values for p-type (n-type) doping along the $x$ and $y$ directions reach $1358.6\,\mu\text{V/K}$ ($1039.1\,\mu\text{V/K}$) and $1353.6\,\mu\text{V/K}$ ($1013.2\,\mu\text{V/K}$), respectively.Due to the trade-off relationship between $S$, the concentration-dependent behavior of $\sigma$ shown in Figure~\ref{fig:13}C and~\ref{fig:13}D exhibits a contrasted trend with that of $S$. Furthermore, the electrical conductivity of the p-type material is significantly lower than that of the n-type, which aligns with the analysis of carrier concentration. Additionally, $\sigma$ increases gradually with temperature.The large $S$ leads to relatively low $\sigma$ in the \ce{PbSe/PbTe} heterostructure monolayer, with values around \SI{1e5}{\siemens\per\meter} at optimal doping concentrations.

\begin{figure}
  \centering
  \includegraphics[angle=0,width=0.88\textwidth]{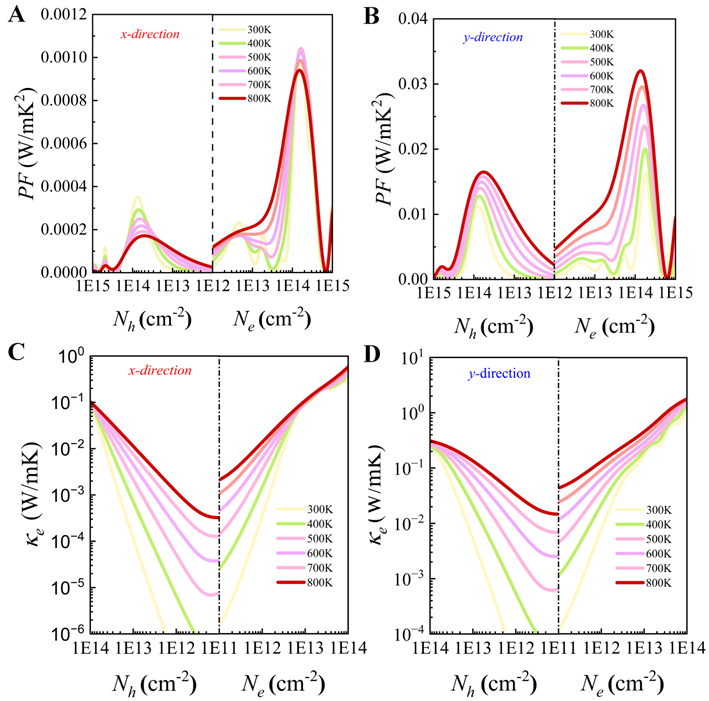}
  \caption{\label{fig:14}  
     Power factor versus carrier concentration of PbSe/PbTe monolayer heterostructure along $x$ and $y$ directions in the temperature range 300 k to 800 K: (A) p-type; (B) n-type. Electronic thermal conductivity ($\kappa_e$) versus carrier concentration: (C) p-type; (D) n-type.}
\end{figure}

Next, the electronic transport performance of the \ce{PbSe/PbTe} monolayer heterostructure is evaluated using the power factor. As shown in Figure~\ref{fig:14}A and~\ref{fig:14}B, the values of power factors under n-type doping are significantly higher than those under p-type doping. Moreover, the power factors along the $y$-direction are evidently greater than those along the $x$-direction. Specifically, under n-type doping, the power factor in the $y$-direction reaches a maximum of \SI{0.035}{\watt\per\meter\per\kelvin\squared}, which is much higher than that of common 2D thermoelectric materials such as \ce{ZrSn2N4} (\SI{3.13}{\milli\watt\per\meter\per\kelvin\squared})~\cite{ref64} and \ce{PbTe} ($\sim$\SI{5.861}{\milli\watt\per\meter\per\kelvin\squared})~\cite{ref65}, indicating the superior electronic transport properties of the \ce{PbSe/PbTe} heterostructure monolayer along the $y$-direction.The observed asymmetry in power factor enhancement between n-type and p-type doping can be attributed to their distinct electronic structures, as previously discussed. The VBM exhibits a relatively flat dispersion, resulting in a larger effective mass for holes, which is favorable for achievinga high Seebeck coefficient. In contrast, the CBM is more dispersive and sharper, leading to a smaller effective mass for electrons and, consequently, higher carrier mobility. This fundamental difference in band curvature underlies the disparity in power factor between n-type and p-type doping.Moreover, the transport parameters listed in Table~\ref{tab:transport_props} quantitatively support this interpretation. The effective mass of n-type carriers is only \SI{0.122}{m_0} in both directions, significantly lower than that of p-type carriers, which ranges from \SI{0.531}{m_0} to \SI{0.627}{m_0}. As a result, the mobility of n-type carriers reaches \SI{1126.8}{\centi\meter\squared\per\volt\per\second} and \SI{1243.0}{\centi\meter\squared\per\volt\per\second} along the $x$ and $y$ directions, respectively—substantially exceeding the mobilities of p-type carriers (\SI{118.87}{\centi\meter\squared\per\volt\per\second} and \SI{240.95}{\centi\meter\squared\per\volt\per\second}). Additionally, the relaxation time ($\tau$) of n-type carriers is generally longer, further enhancing their transport properties.Although the deformation potential constant ($E$) of n-type carriers is slightly higher, the advantages in effective mass and relaxation time compensate for this, yielding superior overall mobility. In summary, the lower effective mass, enhanced mobility, and longer relaxation time of n-type carriers collectively account for the more pronounced power factor enhancement observed under n-type doping.

The electronic thermal conductivity ($\kappa_e$) is calculated according to the Wiedemann-Franz law~\cite{ref66}:

\begin{equation}
\kappa_e = L\sigma T
\label{eq:wiedemann_franz}
\end{equation}

where $L$ is the Lorenz number, $\sigma$ is the electrical conductivity, and $T$ is the temperature. As an important component of the total thermal conductivity, $\kappa_e$ cannot be neglected at high carrier concentrations and may even dominate the heat transport.Figures~\ref{fig:14}C and~\ref{fig:14}D show that the variation trend of $\kappa_e$ with carrier concentration is consistent with that of $\sigma$, and $\kappa_e$ increases with rising temperature. Additionally, the electronic thermal conductivity under n-type doping is significantly higher than that under p-type doping, which may result in lower $ZT$ values for p-type materials. However, due to the overall low electrical conductivity of the \ce{PbSe/PbTe} monolayer heterostructure, its electronic thermal conductivity remains relatively small, below \SI{10}{\watt\per\meter\per\kelvin}, which helps enhance the thermoelectric performance of the material.

\subsection{Thermoelectric Figure of Merit}
\begin{figure}
  \centering
  \includegraphics[angle=0,width=0.88\textwidth]{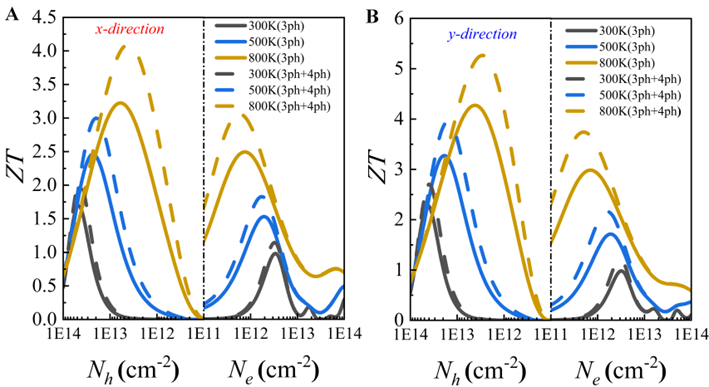}
  \caption{\label{fig:15}  
    Thermoelectric figure of merit ($ZT$) of PbSe/PbTe monolayer heterostructure as a function of carrier concentration along (A) $x$ and (B) $y$ directions at 300 K, 500 K, and 800 K. Solid lines represent results with only three-phonon scattering; dashed lines include four-phonon scattering.}
\end{figure}

Combining its ultralow lattice thermal conductivity, moderate electronic thermal conductivity, and excellent power factor, we further evaluated the thermoelectric figure of merit $ZT$ of the \ce{PbSe/PbTe} monolayer heterostructure. As shown in Figure~\ref{fig:15}, the variation of $ZT$ with carrier concentration at \SI{300}{\kelvin}, \SI{500}{\kelvin}, and \SI{800}{\kelvin} is presented considering both three-phonon (solid lines) and four-phonon (dashed lines) scattering mechanisms.

The results reveal pronounced anisotropic thermoelectric transport behavior, with the $ZT$ values under p-type doping significantly higher than those under n-type doping. This is primarily attributed to the increased electronic thermal conductivity under n-type conditions, which suppresses the enhancement of $ZT$. Moreover, the inclusion of four-phonon scattering mechanisms improves the accuracy of the calculated thermoelectric performance. At \SI{800}{\kelvin}, the maximum $ZT$ values along the $x$ and $y$ directions under p-type doping reach 4.1 and 5.3, respectively---an increase of approximately 23\% compared to the values considering only three-phonon scattering (3.2 and 4.3). For n-type doping, the $ZT$ values also increase from 2.5 ($x$ direction) and 2.9 ($y$ direction) to 3.1 and 3.7, respectively, further confirming the critical role of four-phonon scattering in accurately evaluating thermoelectric performance.

Notably, the optimal $ZT$ value of the \ce{PbSe/PbTe} monolayer heterostructure is 5.3, occurring at \SI{800}{\kelvin} under p-type doping along the $y$-direction, which is significantly higher than the corresponding values for pure monolayer \ce{PbTe} (1.55) and \ce{PbSe} (1.3), highlighting the effectiveness of heterostructure design in substantially enhancing thermoelectric performance. Overall, the \ce{PbSe/PbTe} heterostructure monolayer exhibits outstanding thermoelectric properties at high temperatures, achieving the maximum $ZT$ within the carrier concentration range of \SIrange{e12}{e13}{\per\centi\meter\squared}, demonstrating good experimental feasibility and practical application potential.

Moreover, due to the weak interlayer interactions in the \ce{PbSe/PbTe} monolayer vdW heterostructure, previous studies have indicated the potential for tuning its electronic structure via cross-plane compressive strain. It has been reported that such strain may modify the band structure and band alignment, thereby influencing carrier effective masses and transport properties~\cite{ref67,ref68}. Although this work does not directly investigate the effects of cross-plane compressive strain, based on these studies, we believe this tuning mechanism holds promise for optimizing the electronic performance of \ce{PbSe/PbTe} heterostructures and warrants further exploration in future research.

\section{Conclusions}
Based on first-principles calculations, Boltzmann transport theory, and a four-phonon scattering model assisted by machine-learning interatomic potentials, this work systematically investigates the crystal structural stability, thermal transport, electronic transport, and thermoelectric performance of the two-dimensional \ce{PbSe/PbTe} monolayer heterostructure. The mechanical, dynamical, and thermodynamical stabilities of the heterostructure are verified through calculations of the elastic modulus, phonon dispersion relations, and \textit{ab initio} molecular dynamics simulations.

The \ce{PbSe/PbTe} monolayer heterostructure exhibits an excellent power factor with a peak value reaching \SI{0.035}{\watt\per\meter\per\kelvin\squared}. In terms of thermal transport, owing to the pronounced anharmonicity of the material (Gr\"uneisen parameter $|\gamma|=\num{20}$), relatively low phonon group velocities, and volumetric heat capacity, the lattice thermal conductivity at room temperature reduces to ultralow values of \SI{0.37}{\watt\per\meter\per\kelvin} along the $x$-direction and \SI{0.31}{\watt\per\meter\per\kelvin} along the $y$-direction after including four-phonon scattering, representing decreases of approximately 29\% and 10\%, respectively, compared to considering only three-phonon scattering.

Furthermore, unlike conventional materials where thermal conductivity is mainly dominated by acoustic phonons, optical phonons contribute more than 50\% to $\bm{\kappa}$ in this heterostructure, indicating a unique thermal transport mechanism. Regarding thermoelectric performance, due to the combination of a large power factor and ultralow lattice thermal conductivity, the \ce{PbSe/PbTe} monolayer heterostructure shows significant enhancement in the figure of merit $ZT$ under p-type doping. Notably, at \SI{800}{\kelvin}, the maximum $ZT$ values reach approximately 4.1 and 5.3 along the $x$- and $y$-directions, respectively, substantially outperforming those under n-type doping (approximately 3.1 and 3.7).

In summary, this study not only reveals the outstanding thermoelectric performance and underlying microscopic mechanisms of the \ce{PbSe/PbTe} monolayer heterostructure, highlighting the critical role of higher-order anharmonic scattering in regulating thermal transport, but also provides theoretical guidance and design strategies for enhancing thermoelectric performance via heterostructure engineering.

\section*{Declarations}

\subsection*{Acknowledgments}

None.

\subsection*{Authors’ contributions}
R.T. designed this project under the supervision of K.Z. and Y.-W.F.. R.T. performed the calculations. All authors contribute to the interpretations and writing.

\subsection*{Availability of data and materials}

The data supporting the findings can be found within the manuscript.

\subsection*{Conflicts of interest}

All authors declared that there are no conflicts of interest.


%

\clearpage
\appendix
\section*{Supporting Information}

\setcounter{figure}{0}
\renewcommand{\figurename}{Supplementary Figure}
\setcounter{table}{0}
\renewcommand{\tablename}{Supplementary Table}

\section{Convergence Test Using ShengBTE}
To ensure the accuracy of lattice thermal conductivity calculations, $\mathbf{q}$-point mesh convergence tests were carried out using \textsc{ShengBTE} and \textsc{FourPhonon} for both three-phonon (Supplementary Figure 1) and four-phonon (Supplementary Figure 2) scattering models. The convergence threshold was defined as a variation of less than $0.1\,\mathrm{W/mK}$.

\begin{figure}[H]
    \centering
    \includegraphics[width=0.7\linewidth]{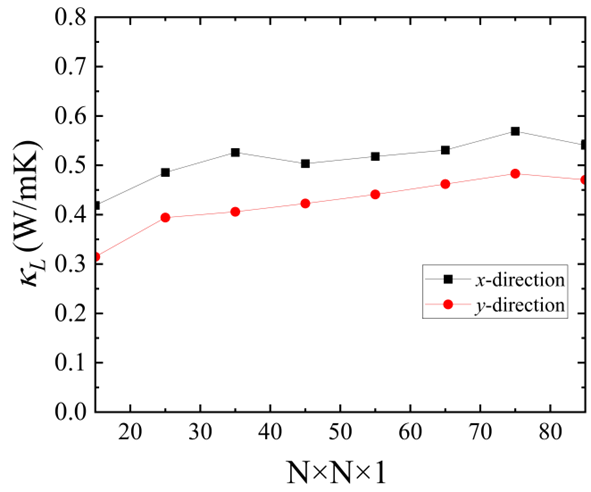}
    \caption{
        Convergence behavior of lattice thermal conductivity in \(\text{RbSe}/\text{RbTe}\) monolayer heterostructure under the three-phonon scattering model, evaluated with varying $\mathbf{q}$-point meshes.
    }
    \label{fig:SI-Fig1}
\end{figure}

\begin{figure}[H]
    \centering
    \includegraphics[width=0.7\linewidth]{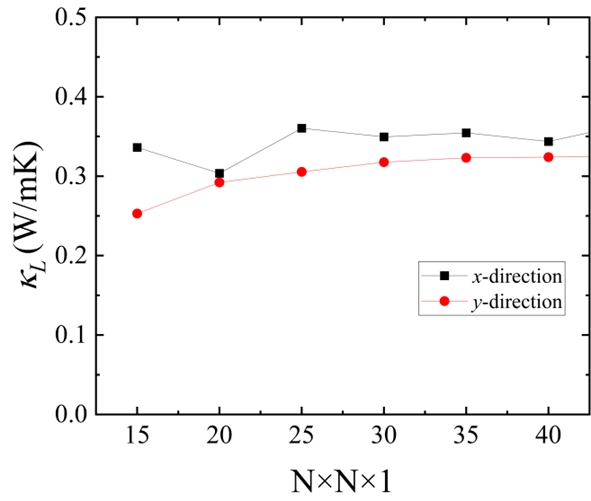}
    \caption{
        Convergence behavior of lattice thermal conductivity in \(\text{RbSe}/\text{RbTe}\) monolayer heterostructure under the Four-phonon scattering model, evaluated with varying $\mathbf{q}$-point meshes.
    }
    \label{fig:SI-Fig2}
\end{figure}
\vfill
\section{Variation of CBM/VBM under Strain}
As shown in Supplementary Figure 3, the electron and hole carrier mobilities were evaluated using the deformation potential theory, considering the strain-induced shifts in the conduction band minimum (CBM) and valence band maximum (VBM).
\begin{figure}[htp]
    \centering
    \includegraphics[width=0.99\linewidth]{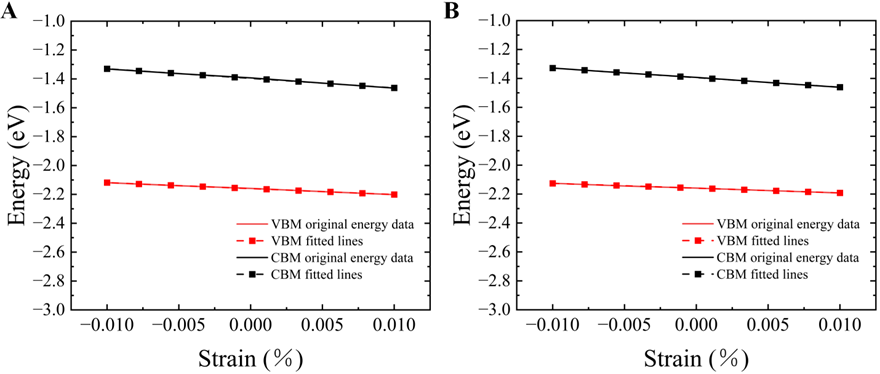}
    \caption{
	    Variation of total energy with uniaxial strain ($-1\%$ to $1\%$) for \ch{PbSe/PbTe} monolayer heterostructure calculated using the PBE functional. The left and right panels correspond to strain applied along the (A) $x$ and (B) $y$ directions, respectively. Red and yellow lines denote the raw energy data and the fitted curves, respectively, for extracting the deformation potential constants.
    }
    \label{fig:SI-Fig3}
\end{figure}

\vfill

\section{Discussion on the Reliability of the MTP Potential}

To verify the applicability of the machine-learned interatomic potential (MLIP) used in this study, namely the Moment Tensor Potential (MTP), for the \ch{RbSe/RbTe} monolayer heterostructure, we present key performance indicators and error statistics from the training process. The MTP model achieved a mean absolute error (MAE) of $0.00040\,\mathrm{eV/atom}$ for atomic energies, with a maximum deviation of $0.00192\,\mathrm{eV/atom}$ and a root mean square error (RMSE) of $0.00059\,\mathrm{eV/atom}$. For atomic forces, the average error was \SI{0.0154}{eV/\angstrom}, with an RMSE of \SI{0.0310}{eV/\angstrom}. The relative error in force fluctuations is only $8.65\%$, well below the commonly accepted threshold of $15\%$ for reliable atomic simulations, indicating good accuracy and predictive capability.\cite{ref67}

To further validate the reliability of the MTP potential in describing the phonon transport properties of the \ch{PbSe/PbTe} monolayer heterostructure, we compared the phonon dispersion curves calculated using MTP with those obtained from density functional theory (DFT). As shown in Supplementary Figure~\ref{suppfig:phonon_dispersion}, the two sets of results are in excellent agreement, demonstrating that the MTP model accurately captures the lattice dynamics of the heterostructure. This confirms the robustness of MTP in predicting phonon transport properties with high fidelity.\cite{ref47}

\begin{figure}[htp]
    \centering
    \includegraphics[width=0.7\linewidth]{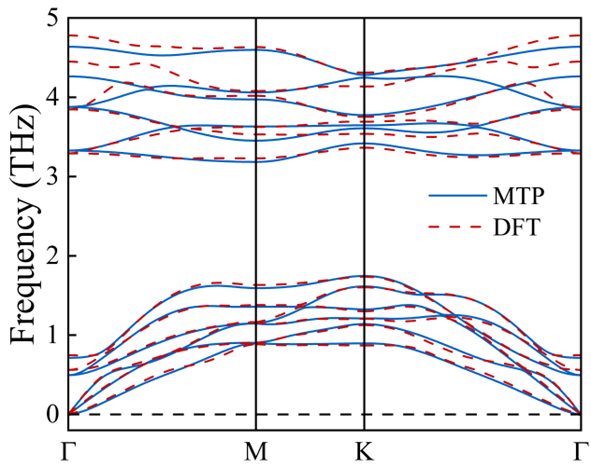} 
    \caption{
        Phonon dispersion curves of \(\text{PbSe}/\text{PbTe}\) monolayer heterostructure calculated using the finite displacement method (red dashed lines) and the machine-learned moment tensor potential (MTP) approach (blue solid lines).
    }
    \label{suppfig:phonon_dispersion}
\end{figure}

\section{Phonon Part}
\begin{figure}
    \centering
    \includegraphics[width=0.7\linewidth]{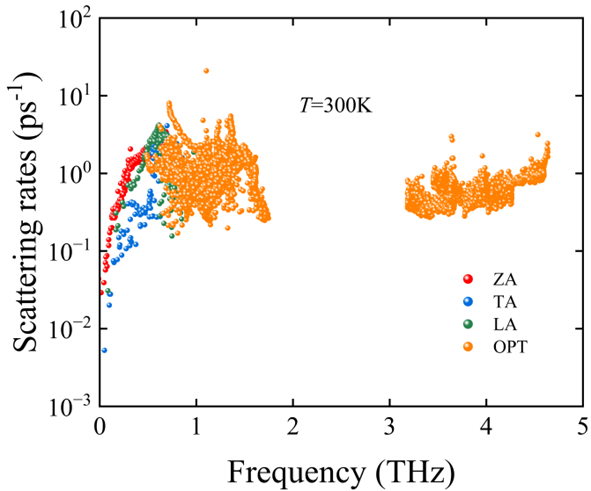} 
    \caption{
        Total phonon scattering rates of each mode as a function of frequency at 300 K under the four-phonon scattering model.}
    \label{suppfig:phonon_scattering}
\end{figure}

\begin{figure}[H]
    \centering
    \includegraphics[width=0.9\linewidth]{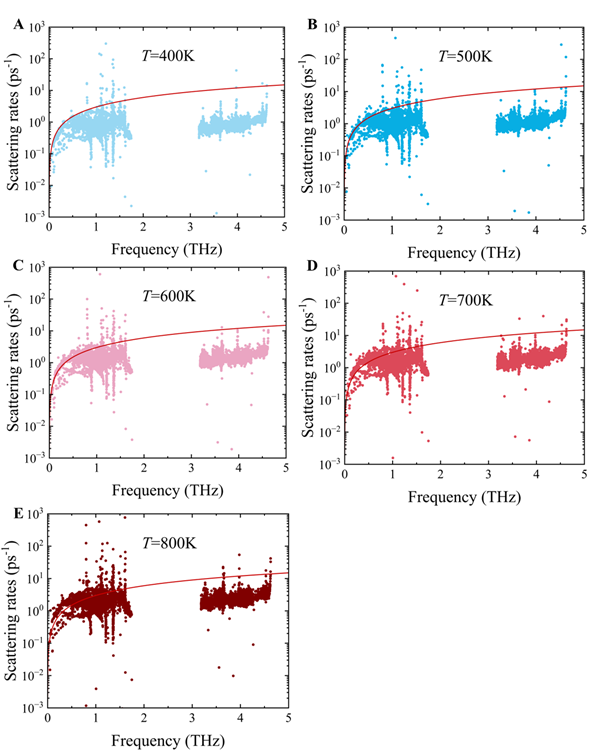}
    \caption{
        Frequency-dependent total scattering rates of the $\mathrm{PbSe}/\mathrm{PbTe}$ monolayer heterostructure calculated under the four-phonon scattering model in the temperature range of 400 K to 800 K. The red line represents $\langle \tau_{\mathrm{ph}}^{-1} \rangle = \omega_{\mathrm{ph}}/(2\pi)$.}
    \label{suppfig:scattering_rates}
\end{figure}

\vfill

\section{Band Gap of pbse/pbte Monolayer Heterostructure}

As shown in Supplementary Table I, the electronic band structures were computed using both the PBE and HSE06 functionals to accurately determine the band gap.

\begin{table}[htp]
\caption{
Band gap values calculated using PBE, PBE+SOC, HSE06, and HSE06+SOC methods for the \ch{PbSe/PbTe} monolayer heterostructure, with comparison to previously reported results.}
\begin{ruledtabular}
\begin{tabular}{lcccc}
\multicolumn{1}{c}{Material} & \multicolumn{4}{c}{Band Gap (eV)} \\
\cline{2-5}
 & PBE & PBE+SOC & HSE06 & HSE06+SOC \\
\hline
\ch{PbSe/PbTe} & 0.77 & 0.53 & 1.25 & 1.05 \\
\end{tabular}
\end{ruledtabular}
\label{tab:band_gaps}
\end{table}

\begin{figure}[htp]
    \centering
    \includegraphics[width=0.7\linewidth]{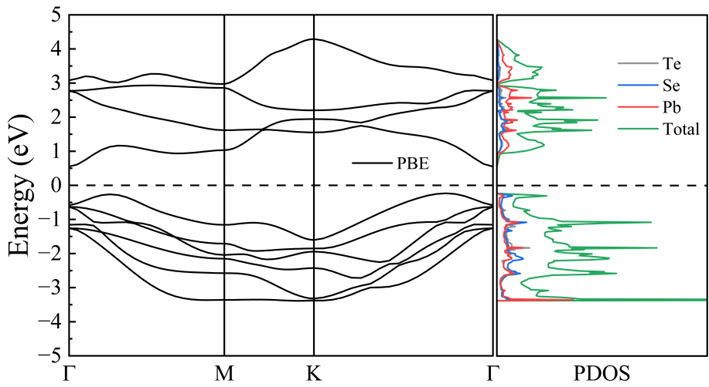}
    \caption{
        Band structures and density of states calculated using PBE methods.}
    \label{suppfig:scattering_rates}
\end{figure}

\clearpage


\end{document}